\documentclass[twocolumn,prd,superscriptaddress,showpacs,preprintnumbers,amsmath,amssymb,nofootinbib]{revtex4}

\pdfoutput=1
\usepackage{epsfig}
\usepackage{subfigure}
\usepackage{amsmath,amssymb}
\usepackage{graphicx}
\usepackage{rotating}
\usepackage{bm}
\usepackage{color}
\usepackage{easybmat}

\def\di{\displaystyle}

\def\bg{\begin{eqnarray}\begin{array}{rcl}\displaystyle}
\def\eg{\end{array} &\di    &\di   \end{eqnarray}}
\def\bm#1{\begin{eqnarray}\begin{array}{#1}\di}
\def\bmo#1{\begin{eqnarray*}\begin{array}{#1}\di}
\def\bml#1#2{\begin{eqnarray}\begin{array}{#1}\label{#2}\di}
\def\bgo{\begin{eqnarray*}\begin{array}{rcl}\displaystyle}
\def\ego{\end{array} &\di    &\di \nonumber  \end{eqnarray*}}
\def\no{\nonumber}
\def\btensor#1#2{\renew\left#1\begin{array}{#2}\di}
\def\brtensor#1#2#3{\ren#3\left#1\begin{array}{#2}}
\def\botensor#1#2{\renew\left#1\begin{array}{#2}}
\def\etensor#1{\end{array}\right#1}

\def\eq#1{(\ref{#1})}
\def\Eq#1{Eq.~(\ref{#1})}

\def\id{1\!\mbox{l}}

\def\ov{\over}
\def\s0#1#2{\mbox{\small{$ \frac{#1}{#2} $}}}
\def\0#1#2{\frac{#1}{#2}}

\def\e{\mbox{\boldmath$\epsilon$}}

\def\CF{{\mathcal F}}

\def\CI{{\mathcal I}}

\def\Z{Z}

\def\ren#1{\renewcommand{\arraystretch}{#1}}

\def\renew{\renewcommand{\arraystretch}{1}}

\definecolor{blue}{rgb}{0,0,1}

\definecolor{green}{rgb}{0,1,0}

\definecolor{red}{rgb}{1,0,0}

\newcommand{\Tr}{\mathrm{Tr}}

\newcommand{\be}{\begin{eqnarray}}
\newcommand{\ee}{\end{eqnarray}}

\begin{document}

\title{The phase diagram of quantum gravity from diffeomorphism-invariant RG-flows}



\author{Ivan Donkin}

\affiliation{Institut f\"ur Theoretische Physik, Universit\"at Heidelberg, 
Philosophenweg 16,
D-69120 Heidelberg, Germany}

\author{Jan M.~Pawlowski}

\affiliation{Institut f\"ur Theoretische Physik, Universit\"at Heidelberg, 
Philosophenweg 16,
D-69120 Heidelberg, Germany}

\affiliation{ExtreMe Matter Institute EMMI, GSI Helmholtzzentrum f\"ur Schwerionenforschung mbH, Planckstr. 1, D-64291 Darmstadt, Germany}

\begin{abstract}
  We evaluate the phase diagram of quantum gravity
  within a fully diffeomorphism-invariant renormalisation group
  approach. The construction is based on the geometrical or
  Vilkovisky-DeWitt effective action. We also resolve the difference
  between the fluctuation metric and the background metric. This
  allows for fully background-independent flows in gravity.

  The results provide further evidence for the ultraviolet fixed point
  scenario in quantum gravity with quantitative changes for the fixed
  point physics. We also find a stable infrared fixed point related to 
  classical Einstein gravity. Implications and possible extensions are discussed.
\end{abstract}

\maketitle

\section{Introduction}\label{sec:intro}

In the past decade the asymptotic safety scenario of quantum gravity
\cite{Weinberg:1980gg} has been explored in quite some
detail. Evidence for a non-trivial UV fixed point (FP) has been
collected with various methods, see e.g.\
\cite{Burgess:2003jk,Niedermaier:2006wt,Hamber:2009mt,Ambjorn:2009ts,Reuter:2007rv,Percacci:2007sz,Litim:2011cp}.
Renormalisation group (RG) approaches to quantum gravity are naturally
well-suited to study such a scenario. In its modern functional form
the renormalisation group is by now very-well developed with many
results in various physics areas, for reviews on gravity and gauge
theories see e.g.\
\cite{Pawlowski:2005xe,Gies:2006wv,Reuter:2007rv,Percacci:2007sz,Rosten:2010vm,Litim:2011cp,Reuter:2012id}.
Since the early works on the functional RG (FRG)
\cite{Reuter:1996cp,Falkenberg:1996bq,Souma:1999at}, which were
carried out within the Einstein-Hilbert approximation, our
understanding of the underlying physics has been extended
tremendously, for reviews see
e.g. \cite{Reuter:2007rv,Percacci:2007sz,Litim:2011cp,Reuter:2012id}. In particular,
the stability of the fixed point scenario has been tested far beyond
the original Einstein-Hilbert truncation. These extensions include
effects generated by the Weyl tensor as well by general terms in the
curvature scalar, e.g.\ \cite{Codello:2007bd,Machado:2007ea}, higher
order derivative terms, e.g.\ \cite{Codello:2006in,Saueressig:2011vn},
ghost fluctuations, e.g.\
\cite{Groh:2010ta,Eichhorn:2009ah,Eichhorn:2010tb}, first attempts on
Lorenzian gravity, e.g.\  \cite{Manrique:2011jc}, as well as the coupling to
matter and gauge fields, e.g.\
\cite{Percacci:2002ie,Daum:2009dn,Harst:2011zx,Eichhorn:2011pc,Folkerts:2011jz}.

The impressive plethora of results, including those obtained in other
approaches, \cite{Burgess:2003jk,Niedermaier:2006wt,
  Percacci:2007sz,Hamber:2009mt,Ambjorn:2009ts,Litim:2011cp}, give us
a firm grip on the asymptotic safety scenario in quantum gravity. This
allows us to study interesting physics related to cosmology and the
dynamics of the full matter-gravity system. Still, all approaches to
quantum gravity have to face the non-trivial task of implementing full
diffeomorphism invariance and reparameterisation invariance of the
theory. This task is tightly linked to the question of background
independence of quantum gravity which is also not fully resolved yet.

In the present work diffeomorphism invariance and background
independence are discussed within the functional RG approach to
gravity. This approach is based on the standard background field
approach to quantum field theory, in which the theory is expanded about a
specific background field configuration. In gravity this is usually
realised within a linear splitting of the full metric $g$ in a
background metric $\bar g$ and a fluctuation $h=g-\bar g$. Finally, the
background is identified with the dynamical metric by setting $h=0$, which
removes the background field dependence, see e.g.\
\cite{Reuter:2008wj}. In this approach the effective
action is invariant under symmetry transformations of the background field
configuration. At its root this is only an auxiliary symmetry whereas
the dynamical symmetry transformations of the fluctuations are 
non-trivially realised.  Note, however, that the fluctuation field $h$
in such an approach has no geometrical meaning, i.e.\ in gravity $h$
is no metric, and in the simpler example of a Yang-Mills theory
the fluctuation field is no connection.

Moreover, the symmetry identities of the fluctuation
fields lead to non-linear relations between fluctuation field Green
functions. It is also possible to derive identities that link
background field Green functions and fluctuation field Green
functions, the Nielsen identities \cite{Nielsen:1975fs}. The Nielsen
identity in combination with the gauge/diffeomorphism covariance of
the background field Green functions provide the non-trivial symmetry
identities of the fluctuation field.  In summary these relations are
chiefly important for the discussion of diffeomorphism invariance as
well as background independence in quantum gravity, and are at the
root of the interpretation of the background correlation functions as
S-matrix elements.

In the present work we put forward a fully diffeomorphism-invariant
FRG approach to quantum gravity,
\cite{Branchina:2003ek,Pawlowski:2003sk, Pawlowski:2005xe,donkin}, by
using the geometrical or Vilkovisky-DeWitt effective action, e.g.\
\cite{Fradkin:1983nw,Vilkovisky:1984st,DeWitt:1988dq,%
  DeWitt:2003pm,Burgess:1987zi,Kunstatter:1991kw}.  Our construction
can be understood as a non-linear upgrade of the standard background
field approach, its linear order giving precisely the background field
relations in the Landau-DeWitt gauge. The gain of such a non-linear
approach is that the fluctuation fields have a geometrical meaning and
can be utilised to compute an effective action which only depends on
the diffeomorphism-invariant part of the fluctuation
fields. Consequently, the geometrical effective action is trivially
diffeomorphism-invariant, and any cutoff procedure applied to these
fluctuation fields maintains diffeomorphism invariance. Still,
fluctuation field Green functions and background metric Green
functions are related to each other by means of a regulator-dependent
Nielsen identity \cite{Pawlowski:2003sk,Pawlowski:2005xe}.

Within this framework, we provide the first fully
diffeomorphism-invariant evaluation of the phase diagram of quantum
gravity including the infrared sector of the theory. Our approach also
allows for a more direct access to the question of background
independence. In a first non-trivial approximation the present work
provides further evidence for the asymptotic safety scenario of
quantum gravity. We also unravel an interesting infrared fixed point
structure. 

In Section~\ref{sec:geo} we briefly recapitulate the geometrical
approach to quantum gravity. Its FRG version as formulated in
\cite{Branchina:2003ek,Pawlowski:2003sk} is introduced in
Section~\ref{sec:geometricalflow}. In Section~\ref{sec:approximation}
we define the approximation which captures the difference between
background metric dependence and fluctuation metric dependence. In
Section~\ref{sec:Nielsen} the Nielsen identity for the regularised
geometrical effective action,
\cite{Pawlowski:2005xe,Pawlowski:2003sk}, is used to derive relations
between different terms in the effective action. In
Section~\ref{sec:backgroundflow} we compute the UV fixed point within
the geometrical approach in the Einstein-Hilbert truncation without
the Nielsen identity. In Section~\ref{sec:dynflow} we utilise the
Nielsen identity to derive both the flow of the background couplings
as well as that of the dynamical couplings. Results on the UV fixed
point scenario within the geometrical approach in the standard
background approximation are presented in
Section~\ref{sec:geobackresults}. In four space-time dimensions they
agree with that obtained in the standard background field approach
within the same background approximation, and in Landau-DeWitt gauge.
In Section~\ref{sec:geoflucresults} we present the results for the
phase diagram of quantum gravity within the fully dynamical
approach. The UV-fixed point scenario agrees qualitatively with that
found in the background field approximation, and compares well with
that found in the bi-metric background field approach put forward in
\cite{Manrique:2010am}, see Section~\ref{subsec:UV}. We also find a
stable infrared fixed point, see Section~\ref{subsec:IR}, and show
that the theory tends towards classical Einstein gravity in the
infrared, see Section~\ref{subsec:physob}. We close with a brief
summary and discussion in Section~\ref{sec:summary}.

\section{Geometrical effective action}\label{sec:geo}
In this Section we briefly review the geometrical approach to quantum
field theory using the notation from
\cite{Pawlowski:2005xe,Pawlowski:2003sk,donkin,DeWitt:2003pm}. The
geometrical approach hinges on the observation that the standard path
integral has no manifest reparameterisation invariance. Put
differently, its standard formulation assumes a flat path integral
measure $d\varphi$ for a given field theory with field
$\varphi$. Neither such a measure nor the related source term $\int_x
J\varphi$ is invariant under field reparameterisations. This apparent
non-invariance can be cured by enhancing the flat measure by an
appropriately defined determinant $\sqrt{\det \gamma}$ of the metric
in field space, $\gamma$, and using a reparameterisation invariant
source term $\int_x J\phi(\bar\varphi,\varphi)$. Here, $\phi$ is
chosen to be a geodesic normal field, i.e. it is the Gaussian normal
coordinate representation of the fluctuating field $\varphi$ with
respect to a chosen background $\bar\varphi$. In linear approximation,
$\phi=\varphi-\bar\varphi$, this reduces to the standard background
field approach.

In gravity the field $\varphi$ is the metric $g$ and the classical
action is the Einstein-Hilbert action $S$, 
\begin{eqnarray}\label{eq:SEH}
&&S[g] = 2 \kappa^2\int d^d x\sqrt{g}\,
\Bigl(-R(g)+2\Lambda\Bigr)\,,
\end{eqnarray}
with curvature scalar $R$ and cosmological constant $\Lambda$. The
prefactor $\kappa^2$ is given by
\begin{equation}\label{eq:kappa}
\kappa^2=\0{1}{32 \pi G_N}\,.
\end{equation}
where $G_N$ is the Newton constant. The basic object in the geometrical
approach to gravity is the configuration space of the theory, $\Phi =
\{g_{\mu\nu}\}$, equipped with the natural action of the
diffeomorphism group $\,\mathcal{G}$. There is a one-parameter family
of ultralocal group-invariant supermetrics on $\Phi$
\begin{eqnarray}\label{eq:supermetric}
  &\gamma^{\mu\nu\rho'\sigma'}(x,x')= \Big[\,\frac{1}{2}\,
  g^{\mu\rho'}(x) g^{\nu\sigma'}(x) + \frac{1}{2}g^{\mu\sigma'}(x)
  g^{\nu\rho'}(x) \nonumber\\[1ex] 
  &-\theta\, g^{\mu\nu}(x) g^{\rho'\sigma'}(x) \,\Big] \sqrt{g(x)}\,\sqrt{g(x')}
  \,\delta(x,x')  
\end{eqnarray}
labelled by a continuous real parameter $\theta$. For the remainder of
the paper we fix $\theta = -1$.

\begin{figure}[h]
\includegraphics[width=1\columnwidth]{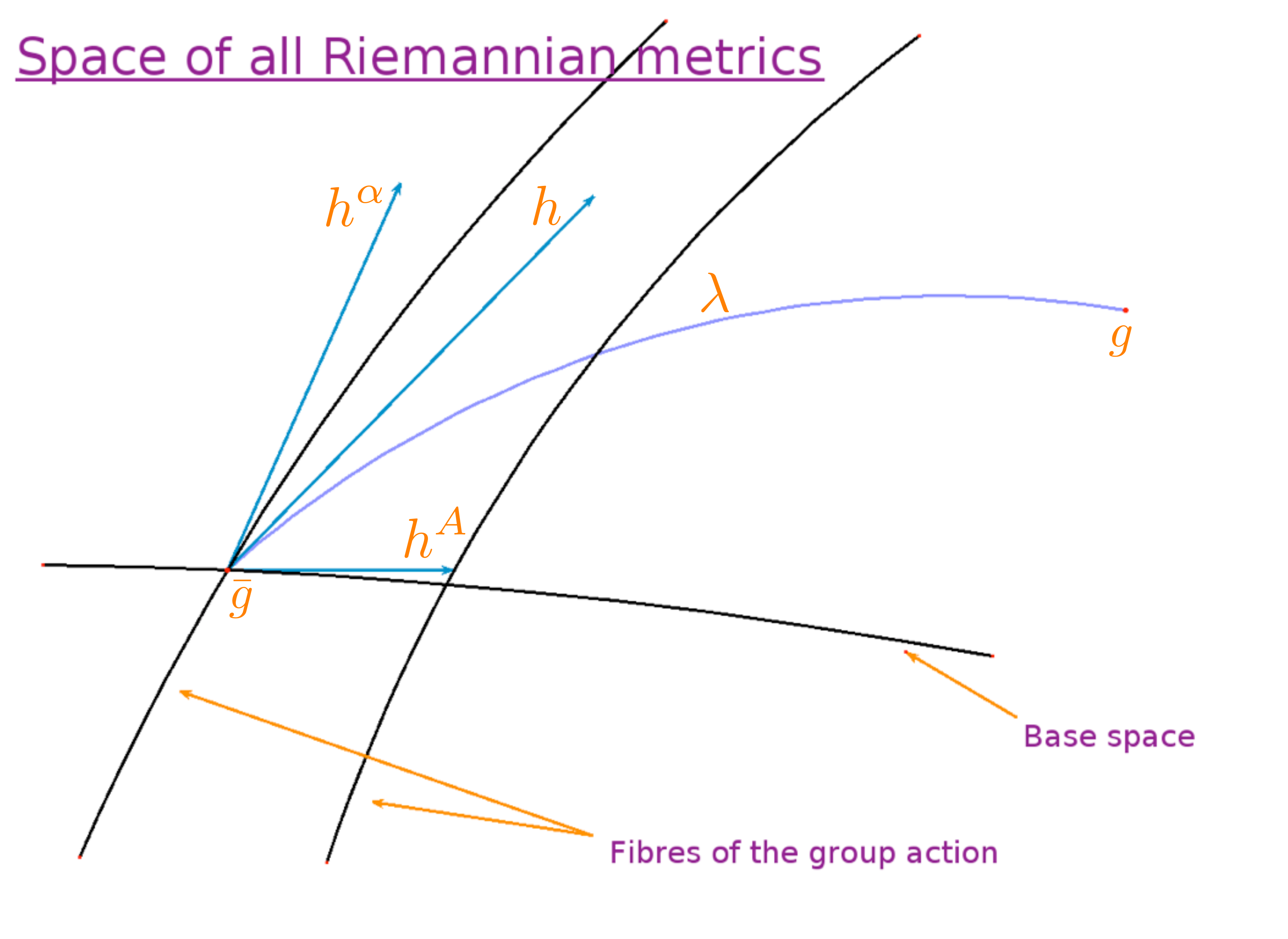}
  \caption{Geodesic w.r.t.\ the Vilkovisky connection from $\bar g$
    to $g$. $\sigma$ is the the tangent vector at $\bar g$ on this
    geodesic, $h^A$ is the diffeomorphism-invariant projection,
    and $h^\alpha$ the projection on the diffeomorphism fibre.}
\label{fig:geometry}\end{figure}

In the standard background field approach one expands the metric $g$
about a given background metric $\bar g$ within a linear split, $g=\bar g+h$
with fluctuation field $h$. Such a parameterisation entails that the
fluctuation $h$ is neither a metric nor a vector, i.e. it has no
geometrical meaning. In turn, within the geometrical approach we
define $h$ as a tangent vector at $\bar g$ and $\sigma^a[\bar g;g]=-h^a$ as the
geodesic normal coordinate of $g$ with respect to $\bar{g}$, see e.g.\ \cite{DeWitt:2003pm},  
\begin{eqnarray}\label{eq:h} 
\sigma^a[\bar g;g]= (\bar s-s){d\lambda^a\ov d s}(\bar s) 
\end{eqnarray}
This construction is illustrated in Figure~\ref{fig:geometry}. 
The geodesics $\lambda(s)$ are taken with respect to  
Vilkovisky's connection and satisfy $\lambda(\bar s) = \bar g$ and 
$\lambda(s) = g$, and $s$ is the affine parameter of the
geodesic.  Heuristically speaking, Vilkovisky's connection is
designed to maximally disentangle the fibre from the base space. It is
defined through its Christoffel symbols
\begin{eqnarray}\label{eq:Vcon}
\Gamma_V{}^i{}_{jk}=\Gamma_\gamma{}^i{}_{jk}- Q^i{}_{\alpha\cdot(j}\,
\omega^\alpha{}_{k)}+\s012 \omega^\alpha{}_{(j}Q^i{}_{\alpha\cdot l}
Q^l{}_{\beta}\,\omega^\beta{}_{k)}
\end{eqnarray}
where $\Gamma_\gamma{}$ are the Christoffel symbols of the Riemannian connection 
induced by the supermetric $\gamma$ and $Q_{\alpha}$ are the generators of the 
diffeomorphism group. The $\omega^\alpha{}_{i}$ are the components of
the unique connection one-form $\bf{\omega}^\alpha$ determined by 
$\gamma$ and $Q_{\alpha}$, i.e.
\begin{equation}
\omega^\alpha = \mathfrak{G}^{\alpha\beta} \, \gamma(Q_{\beta},\,\cdot\,\,)
\end{equation}
where $\,\mathfrak{G}^{\alpha\beta}\,$ stands for the inverse operator
of $\,\,-\,\gamma(Q_{\alpha}, Q_{\beta})$. With DeWitt's condensed
notation the index $i = (x,\mu)$ labels space-time $x$ and Lorentz
indices $\mu$. Additionally, the subscripts $\,\cdot j\,$ denote
covariant derivatives with respect to $\Gamma_\gamma{}$ and the
parenthesis in the subscripts indicate symmetrization of the indices
embraced. More details in the context of functional RG flows in the
geometrical approach can be found in
\cite{Pawlowski:2005xe,Pawlowski:2003sk,donkin,Branchina:2003ek}.

The above geometrical construction allows us to define the path
integral of the theory in a manifestly reparameterisation
invariant way
\begin{eqnarray}\label{eq:geoeff}
e^{-\hat\Gamma[\bar g; h]} & =&\int \mathcal{D} g\,\sqrt{\det \gamma} \,\,\delta(\mathcal{F} )
\,\det M [\bar{g};g]\, \nonumber\\
&& \!\!\times \exp\left\{\,-S[g] \,- \int \0{\delta\hat\Gamma}{\delta h}\cdot 
(\sigma[\bar g;g]+ h)\right\} \,.
\end{eqnarray}
Here $\,\mathcal{D} g\sqrt{\det \gamma}\,$ stands for the volume form on $\Phi$,
$\mathcal{F}=0$ is the gauge fixing condition and $\,\det M\,$is the
determinant of the ghost operator which depends both on the background
$\bar{g}$ and the fluctuating field. We emphasise that the gauge
fixing is only introduce for the sake of convenience: the effective
action $\hat \Gamma$ does not depend on it. This goes hand in hand
with the fact that the geometrical effective action $\hat\Gamma$ in
\eq{eq:geoeff} only depends on the diffeomorphism-invariant part $h^A$ of
the field $h$ with coordinates $\,h^a = (h^A, h^\alpha)$,
\begin{equation}\label{eq:hA}
h^a= - \sigma^a[\bar g;g]\,,\quad h^A=(\Pi h)^A\,, \quad h^\alpha = 
\left((\id -\Pi) h\right)^\alpha\,.  
\end{equation}  
Here we have introduced the horizontal projection operator
$\Pi=\Pi(\bar g)$ on the diffeomorphism-invariant part of $h$, see
e.g.~\cite{Pawlowski:2003sk,Pawlowski:2005xe,DeWitt:2003pm}. The part of the
geodesic normal field tangential to the  fibre, $h^\alpha$, 
drops out. Note also, that in the linear approximation,
$h$ is equivalent with the metric fluctuation $g-\bar g$ in the
background field approach, $h=g-\bar g+O(h^2)$. With these
prerequisites it is possible to rewrite the path integral in
\eq{eq:geoeff} in terms of the field $h$
\begin{eqnarray}\nonumber 
  e^{-\Gamma[\bar g; h]}&= &\int \mathcal{D} \hat h^A\,\sqrt{\det 
    \gamma^{AB}}   \, e^{-S[\bar g;\hat h]}  \int \mathcal{D} \hat h^\alpha \sqrt{
    \det \gamma^{\alpha\beta}}  \\[1ex]
  && \hspace{-.2cm}\times \,\,e^{-S_{\rm gf}[\bar g; h]} \,{\det}^{1/2}\CF^2 \det M\, e^{\int 
    \0{\delta\Gamma}{\delta h}\cdot (\hat h-h)}\,,\label{eq:geoeffh}\end{eqnarray}
where $\,\mathcal{F}^2\,$ is the distribution kernel of the gauge fixing term $\,S_{\rm gf}$, 
\begin{eqnarray}\label{eq:Sgf}
S_{\rm gf}[\bar g; h]= \frac{\kappa^2}{\alpha}\int d^d x \sqrt{\bar{g}} \,\bar{g}^{\mu\nu}
\CF_{\mu} \CF_{\nu}\,,
\end{eqnarray}
with $\kappa$ defined as in \eq{eq:kappa}. 
We emphasise that the Gaussian integration over the 
fibre field $\hat h^\alpha$ in \eq{eq:geoeffh} is only kept for the sake of convenience. 
Performing it would make explicit that the effective action $\Gamma[\bar g; h]$
defined in \eq{eq:geoeffh} only depends on $h^A$ up to 
the gauge fixing term, 
\begin{eqnarray}\label{eq:trivialsplit}
\Gamma[\bar g; h]=\hat\Gamma[\bar g; h^A] + S_{\rm gf}[\bar g;h]\,. 
\end{eqnarray}
In the following we impose a linear gauge fixing condition 
\begin{equation}\label{eq:gauge}
\CF_{\tau} =   F_\tau^{\mu\nu}[\bar g] \,\hat h_{\mu\nu}=0\,.
\end{equation} 
with a linear operator $F_\tau^{\mu\nu}[\bar g]$ which depends on the
background $\bar{g}$. In this case the ghost operator $\,M\,$ in \eq{eq:geoeffh} 
depends solely on the background field configuration $\bar{g}$ but not on 
the fluctuating field $\hat h$. Put intuitively,  $\,\det M\,$ 
accounts for the fact that the gauge fixing surface specified by 
$\,\mathcal{F}\,$ intersects each gauge orbit at a different angle. In
general the intersection angle will depend on the dynamical field configuration
$g$ parametrizing the orbit, see \eq{eq:geoeff}. Note, however, that 
the path integral in \eq{eq:geoeffh} is taken over a linear manifold  
and the gauge orbits are linear hypersurfaces -- the vertical subspaces
spanned by the $h^\alpha$ coordinates. Then, with \eq{eq:gauge} it is clear 
that the gauge fixing surface will intersect each orbit at the same angle 
leading to a constant $\,M$.

It is convenient though not necessary to choose the gauge fixing such that 
$h^A$ satisfies \eq{eq:gauge}, see e.g.~\cite{Pawlowski:2003sk}. With 
\eq{eq:hA} this amounts to $F\cdot  \Pi=0$. Then it is evident that 
the Gaussian integration over $h^\alpha$ drops out, leading to 
purely background field dependent terms multiplied by $\det
\gamma^{\alpha\beta}$. Nonetheless it turns out to be
convenient to keep the gauge fixing term and use it in order to
facilitate computations. Specifically we choose
\begin{equation}\label{eq:gaugefix}
F_\tau^{\mu\nu}[\bar g] = \gamma^{\mu\nu\rho'\sigma'}[\bar{g}]\, 
Q_{\rho'\sigma',\,\tau} [\bar{g}]
\end{equation}
Using the well-known expression for the generators of the diffeomorphism
group
\begin{equation}\label{eq:generators}
Q^i{}_\alpha [\bar g]= Q_\tau^{\rho'\sigma'} [\bar g]= -\,\delta^{\rho'}_\tau 
\bar{\nabla}^{\sigma'} -\delta^{\sigma'}_\tau \bar{\nabla}^{\rho'}
\end{equation}
we obtain
\begin{equation}\label{eq:gaugefixing}
F_\tau^{\mu\nu}[\bar g] =  -2 \left( \,\delta_\tau^{(\mu} \bar{\nabla}^{\nu)}
+ \frac{1}{2}\,\theta \,\bar{g}^{\mu\nu} \bar{\nabla}_{\tau}\, \right) \,.
\end{equation}
Here, $\,\bar{\nabla}\,$ is the covariant derivative with respect to the background 
metric connection. It is now straightforward to show that
\begin{eqnarray}\label{eq:gaugefixaction} 
S_{\rm gf} &=& -\frac{\kappa^2}{\alpha}\int d^d x\,\sqrt{\bar{g}}\,\hat h_{\mu\nu}
\bigg( \bar{g}^{\mu\sigma'}\bar{\nabla}^{\nu}\bar{\nabla}^{\rho'} \nonumber\\
&&+ \theta\,\bar{g}^{\mu\nu}\bar{\nabla}^{\rho'}\bar{\nabla}^{\sigma'} + 
\frac{1}{4}\,\theta^2\,\bar{g}^{\mu\nu}\bar{g}^{\rho'\sigma'}\bar{\Delta}\bigg)
\hat h_{\rho'\sigma'} \,,
\end{eqnarray}
with $\alpha$ being the gauge-fixing parameter and $\bar{\Delta}
\equiv \Delta_{\bar{g}}$ the Laplace operator constructed from the
background metric. Finally we discuss the $\sqrt{\det \gamma}$-terms
in \eq{eq:geoeffh}, for details see e.g.\ \cite{DeWitt:2003pm}. First of all 
we note that the full metric $\gamma$ does not depend on the $h^\alpha$ due to the
vanishing Lie-derivative $\mathcal{L}_{Q_\alpha} \gamma = 0$. The
horizontal part $\gamma^{AB}$ does not depend on the $h^A$ either which
leaves $\gamma^{\alpha\beta}$ as the only dynamical object. Explicitly it reads  
\begin{eqnarray}\label{eq:gammafibre}
\gamma^{\alpha\beta} = \Big(\,2\delta^{\mu}_{\nu}\Delta +
2R^{\mu}_{\nu} + 2(1 + \theta)\nabla^{\mu}\nabla_{\nu}\, \Big)\, \delta(x, x')
\end{eqnarray}
with $\,\alpha =(\mu, x)\,$ and $\,\beta = (\nu,
x')\,$. The determinant of $\gamma^{\alpha\beta}$ can be rewritten in
terms of a Grassmann integral,
\begin{eqnarray}\label{eq:Berezin}
&&\hspace{-.8cm}\det \gamma^{\alpha\beta}[\,\bar{g}, h^A]\\\nonumber 
&=& \,\int D\bar{c} Dc\,
e^{\int d^4x\sqrt{g}\,\bar{c}_{\mu} \Big(\,2\delta^{\mu}_{\nu}\Delta +
2R^{\mu}_{\nu} + 2(1 + \theta)\nabla^{\mu}\nabla_{\nu}\, \Big) \,c^{\nu} }\,.
\end{eqnarray}
Here $\,c(x)\,$ and $\,\bar{c}(x)\,$ are anti-commuting Grassmann
fields and $g$ is a metric with $\,-\,\sigma^A[\bar{g},g] =
h^A$. Eq.\eq{eq:Berezin} can most easily be understood as the
geometric analogue of the usual ghost action. This leaves us with the
final expression for the geometrical effective action,
\begin{eqnarray}\label{eq:finalGamma} 
  &&\,\,e^{-\Gamma[\bar g; h]}= \int \mathcal{D} \hat h^A\,\sqrt{\det 
    \gamma^{AB}} \int \mathcal{D} \hat h^\alpha  \,\int \mathcal{D} \bar{c} \,\mathcal{D} c 
  \quad \nonumber \\[1ex]
  && \hspace{-.2cm}\times\,e^{-S[\bar g;\hat h, c ,\bar c]}  \,{\det}^{1/2}\CF^2 \det M\, e^{\int 
    \0{\delta\Gamma}{\delta h}\cdot (\hat h-h)}\,\,,
\end{eqnarray}
where the remaining measure factors of the path integral only lead to
background-dependent terms and $S[\bar g;\hat h, c,\bar c]$ is the
full gauge-fixed action,
\begin{eqnarray}\label{eq:finalS}
&&\hspace{-.7cm}S[\bar g;\hat h, c,\bar c] =2 \kappa^2 \int d^d x\sqrt{g}\,
  \left(-R(g)+2\bar \Lambda_k\right)+S_{\rm gf}\nonumber\\
 && \hspace{-.5cm}+2\int d^d x\sqrt{g}\,\bar{C}_{\mu} 
  \Big(\delta^{\mu}_{\nu}\Delta+
  R^{\mu}_{\nu}+(1+\theta)\nabla^{\mu}\nabla_{\nu}\Big) C^{\nu}\,.
\end{eqnarray}
We emphasise again that even though the path integral
\eq{eq:finalGamma} is defined similarly to the standard gauge-fixed
approach, the effective action $\,\Gamma-S_{\rm gf}$ does not depend
on the gauge fixing.

\section{Geometrical RG-flows} \label{sec:geometricalflow}

The geometrical approach put forward in the last section
allows for a diffeomorphism-invariant infrared regularisation as the
dynamical field $h^A$ is diffeomorphism-invariant. Flow equations for the
geometrical effective action have first been put forward in
\cite{Branchina:2003ek} for the sharp-cutoff and in
\cite{Pawlowski:2003sk} for general regulators. The approach has been
put to work in the Einstein-Hilbert approximation in \cite{donkin}. In
\cite{Pawlowski:2003sk} it has been shown that, despite manifest
diffeomorphism or gauge invariance, the approach is subject to
non-trivial, regulator-dependent Nielsen identities. Heuristically 
speaking, these identities carry the information about the unitarity
of the theory. This interesting and important relation will be discussed elsewhere. 

A diffeomorphism-invariant infrared regularisation can now be applied
to the theory by modifying the propagation of the fluctuation fields
through the substitution $S[\bar g;\hat h, c,\bar c]\to S[\bar
g;\hat h, c,\bar c]+ \Delta S_k[\bar g;\hat h, c,\bar c]$ with the
cut-off term
\begin{eqnarray}\nonumber
\Delta S_k[\bar g; \hat h,\bar c,c]&=& \012 \int  d^4 x \sqrt{\bar g} \,\hat h_{\mu\nu} 
\mathcal{R}^{\mu\nu\rho\sigma}_k[\bar g] \,\hat h_{\rho\sigma} \\[1ex]  
&& + \int  d^4 x \sqrt{\bar g} \,\bar c_{\mu}\, \mathcal{R}^{\mu\nu}_k[\bar g] \,c_{\nu}\, .
\label{eq:regterm} \end{eqnarray} 
Note that the regulators $\,\mathcal{R}^{\mu\nu\rho\sigma}_k\,$ and $\,\mathcal{R}^{\mu\nu}_k\,$ 
only depend on the background field configuration. For convenience, we further demand that 
$\,\mathcal{R}^{\mu\nu\rho\sigma}_k\,$ should satisfy $\mathcal{R}_{k,\alpha A}=\mathcal{R}_{k,A\alpha}=0$, 
see \cite{Pawlowski:2003sk}. This disentangles the trivial flow of the $h^\alpha$-part of the action 
from the dynamical flow of the $h^A$-part. Inserting the regulator terms into the path integral 
\eq{eq:finalGamma} we are led to the Wetterich equation for quantum gravity within the geometrical approach, 
\begin{equation}\label{eq:flow}
\partial_t \Gamma_k[\bar g; \phi]=\012 
\Tr \0{1}{\Gamma_k^{(2)}[\bar g; \phi]+
\mathcal{R}_k[\bar g] } \partial_t \mathcal{R}_k[\bar g]\,,  
\end{equation}
where the trace sums over momenta, internal indices and all field
species with a relative minus sign for Grassmann fields. The
super-field $\phi=(h,C,\bar C)$ contains all fluctuation fields. The
components are the expectation values of the dynamical fields, i.e.\ 
\begin{equation}\label{eq:EVs}
  h=\langle\hat h\rangle\,\quad C_{\mu}=
\langle c_{\mu}\rangle\,,\quad \bar C_{\mu}=\langle \bar
  c_{\mu}\rangle\,.
\end{equation} 
and the two-point function $\Gamma_k^{(2)}[\bar g; \phi]$ is 
the second derivative of the effective action $\Gamma$ w.r.t.\ the field $\phi$,  
\begin{equation}\label{eq:Gamma2}
\Gamma_k^{(2)}[\bar g; \phi]=\0{\delta^2\Gamma_k}{\delta\phi_i\delta\phi_j}\,. 
\end{equation}
The regulator $\mathcal{R}_k$ is diagonal in superfield space with
diagonal components
\begin{equation}\label{eq:Rdiag}
\mathcal{R}_{k,hh}=(\mathcal{R}_k^{\mu\nu\rho\sigma})\,,\qquad 
\mathcal{R}_{k,C\bar C}=(\mathcal{R}_k^{\mu\nu})=-(\mathcal{R}_{k}^{\nu\mu})\,.
\end{equation}
As can be immediately inferred from the
construction of the geometrical effective action, we also have in
general
\begin{equation}\label{eq:dGhat=dG}
  \0{\delta\partial_t \Gamma_k[\bar g; \phi]}{\delta\phi}=
  \0{\delta \partial_t \hat\Gamma_k[\bar g; \phi]}{\delta\phi}\,, 
\end{equation}
i.e. the flows of $\hat \Gamma_k$ and $\Gamma_k$ agree up to
normalisation factors that might depend on the background metric. In
particular this entails that the gauge fixing term does not
flow. Ultimately we are interested in the evolution of
$\,\Gamma_k[\bar g;0]$ with the cut-off scale $k$. In this
case the propagator on the right hand side of \eq{eq:flow} can be
rewritten as 
\begin{equation}\label{eq:covariantgeom}
  \Gamma_k^{(2)}[\bar g; h=0] =  \nabla^{2}\hat\Gamma_k[ \bar g, g
  = \bar g]+S_{\rm gf}^{(2)} \,\,. 
\end{equation}
In \eq{eq:covariantgeom} we have used that $h= - \sigma[\bar g;g]$. The
second covariant derivate $\nabla^{2}$, taken with respect to
$\Gamma^V$, acts on the full dynamical metric field $g$. For
notational convenience we omitted the ghost fields. Note that within
the standard background field approach \eq{eq:covariantgeom} simply
reads
\begin{equation}\label{eq:covariantbground}
\Gamma_k^{(2)}[\bar g; h=0]  = \0{\Gamma_k[\bar g; g= \bar g ]}{\delta g^2}\,.
\end{equation} 
The symmetric tensor $h$ can be further decomposed with the York
transverse-traceless decomposition valid for spherical background
geometries. This decomposition together with that of the ghosts is
detailed in Appendix~\ref{app:york}. In the present work all diagonal
modes $\phi_i$ are regularised with regulators
\begin{equation}\label{eq:regulators} 
  \mathcal{R}_{k,i} = {\cal T}_i\, k^2 r(x)\,,\qquad {\rm with} 
  \qquad x=-\0{\Delta_{\bar g}}{k^2}\,.
\end{equation} 
where $r(x)$ is a dimensional shape function and the prefactor
${\cal T}_i$ accounts for the tensorial structure of the respective
mode. The complete list of regulators can be found in
Appendix~\ref{app:regulators}.

We close this section with a discussion of the practical
implementation of the flow \eq{eq:flow} in the graviton sector in a
given approximation. This repeats the discussion concerning the
trivial difference between the diffeomorphism-invariant effective
action, here $\hat\Gamma_k$, and the trivially gauge-fixed effective
action, here $\Gamma_k=\hat\Gamma_k+S_{\rm gf}$, in the context of the
flow equation. The general derivations are done in detail in
\cite{Pawlowski:2003sk}. 

Approximations or parameterisations of the effective action $\Gamma_k$
contain a diffeomorphism-invariant functional of $g$ such as the
Einstein-Hilbert action. This functional has to be accompanied by
terms which preserve symmetry constraints such as the Nielsen
identities. The flow depends on the second derivative of $\Gamma_k$
w.r.t. $h$ which has to be extracted from the action. For functionals
of $g$ this amounts to taking second derivatives w.r.t.  Vilkovisky's
connection. Here we discuss how this task can be reduced to computing
Riemannian covariant derivatives at $g=\bar g$. Separating the
graviton and ghost contributions and identifying $g=\bar g$, the flow
reads with \eq{eq:trivialsplit}
\begin{eqnarray}\nonumber 
\partial_t \Gamma_{k}&=&\frac{1}{2} 
\Tr \frac{1}{\,\nabla^2 \,\hat \Gamma_{k}+S_{\rm gf}^{(2)}+
\mathcal{R}_k^{\rm grav} } \,\partial_t \mathcal{R}^{\rm grav}_k\\
&& +\,\,{\rm ghost-contr.} 
\label{eq:dipl}\end{eqnarray}
Since $\hat \Gamma_{k}$ is a diffeomorphism-invariant functional, the
covariant derivative $\hat \Gamma_{k}^{(2)}=\nabla^2 \,\Gamma_{k}$
has only a transversal part. If we choose a purely transversal regulator
$\mathcal{R}^{\rm grav}_k[\bar g]$, i.e.
\begin{equation}
\mathcal{R}^{\rm grav}_k[\bar g] =
\left(\begin{BMAT}(rc){c:c}{c:c}
\mathcal{R}^{\rm grav}_{k,\perp}[\bar g] \,\,\, & \,\,\,0 \,\,\,\\
\medskip
0 & \,\,\,0 \,\,\,
\end{BMAT}
\right)\,,
\end{equation}
the term $S_{\rm gf}^{(2)}$ drops out from the flow. This entails that
the geometrical flow cannot depend on the gauge fixing, for the
general argument see \cite{Pawlowski:2003sk}. Thus, we could as well
work solely with $\hat \Gamma_k$ and its propagator. However, for
practical computations it turns out to be more convenient to invert
the propagator on the full transversal + longitudinal space, hence
using $\Gamma_k$ instead of $\hat \Gamma_k$. It is also here where we
make use of the crucial identity (\ref{eq:approxprop}). Schematically
we have at $g=\bar g$,
\begin{equation}
\nabla^2 \hat \Gamma_{k} =\Pi[\bar{g}]\cdot 
 \nabla^2_{\gamma} \,\hat\Gamma_{k}[\bar{g}] \cdot\Pi[\bar{g}] 
\end{equation}
see \eq{eq:hA}. With $F \cdot \Pi [\bar{g}]=0$ and \eq{eq:gauge} the
gauge-fixing term has the form
\begin{equation}\label{eq:S2gf}
S^{(2)}_{\rm gf}[\bar g] \, = -\0{\kappa^2}{\alpha}
\left(\begin{BMAT}(rc){c:c}{c:c}
\,\,\, 0 \,\,\, & 0 \\
\,\,\, 0 \,\,\, & \,\,\,{\cal F}\cdot {\cal F}\,\,\,
\end{BMAT}
\right)\,\,,
\end{equation}
It is proportional to $1/\alpha$ and diverges for $\alpha\to 0$.  
We also introduce a corresponding longitudinal part to the regulator 
\begin{equation}\label{eq:Rtrans+Rlong}
\mathcal{R}_{k}^{\rm grav}[\bar g] =
\left(\begin{BMAT}(rc){c:c}{c:c}
\,\,\,\mathcal{R}^{\rm grav}_{k,\perp}[\bar g]\,\,\, &0\\
\medskip
0 & \,\,\,\mathcal{R}_{k,\rm L}^{\rm grav}[\bar g]\,\,\,
\end{BMAT}
\right)\,\,.
\end{equation}
This modification adds a trivial $\bar g$-dependent part to the flow,
see also \eq{eq:dGhat=dG}. Note also that even though $\Gamma_{k}$ is
a diffeomorphism-invariant functional, its covariant derivative with
respect to the metric connection is not. This is taken into account by
writing schematically
\begin{equation}
\nabla^2_{\gamma} \, \hat \Gamma_{k} =
\left(\begin{BMAT}(rc){c:c}{c:c}
\,\,\,\,\,\nabla^2 \, \hat\Gamma_{k}  \,\,\,\,\, & \,\,\,B \,\,\,\\
\medskip
\,\,\,\,\, C \,\,\,\,\, & \,\,\,D \,\,\,
\end{BMAT}
\right)\,\,
\end{equation}
where the subscript $\,\gamma\,$ indicates that the covariant
derivative is taken with respect to the metric rather than the
Vilkovisky connection. Now we consider 
\begin{eqnarray}
& & \nabla^2_{\gamma} \, \hat\Gamma_{k} +S^{(2)}_{\rm gf}+ 
\mathcal{R}_{k}^{\rm grav} \\ \nonumber  &=&  
\left(\begin{BMAT}(rc){c:c}{c:c}
\,\,\,\,\,\nabla^2 \, \hat \Gamma_{k} +
\mathcal{R}^{\rm grav}_{k,\perp}[\bar g]  \,\,\,\,\, & \,\,\,B \,\,\,\\
\medskip
\,\,\,\,\, C \,\,\,\,\, & \,D +  \mathcal{R}_{k,\rm L}^{\rm grav}
 + S^{(2)}_{\rm gf}\,\,\,
\end{BMAT}
\right)\,,
\end{eqnarray}
in a slight abuse of notation, where $T$ is completely longitudinal.
This applies to the gauge fixing term, $S^{(2)}_{\rm gf}$ in
\eq{eq:S2gf}, and to a sum of gauge fixing term and longitudinal
regulator as introduced in \eq{eq:Rtrans+Rlong}. In these cases we have
\begin{eqnarray}
&& \lim_{\alpha \to 0} 
\left(\begin{BMAT}(rc){c:c}{c:c}
\nabla^2 \hat \Gamma_{k} +\mathcal{R}^{\rm grav}_{k,\perp} & B\\
\medskip
C &  D +  \mathcal{R}_{k,\rm L}^{\rm grav} + S^{(2)}_{\rm gf}
\end{BMAT}
\right)^{-1} \nonumber\\
&=& \left(\begin{BMAT}(rc){c:c}{c:c}
\,\,\,\,\,\Pi\cdot\0{1}{\nabla^2 \, \hat \Gamma_{k} 
+\mathcal{R}^{\rm grav}_{k,\perp}} \cdot\Pi\,\,\,\,\, & \,\,\,0\,\,\, \\
\medskip
0 & \,\,\,0 \,\,\,
\end{BMAT}
\right) 
\end{eqnarray}
where, as before, the $\pi$'s indicate that we are inverting on the
transversal subspace. In summary we can rewrite the right hand side of
eq.(\ref{eq:flow}) at $g=\bar g$ in the form 
\begin{eqnarray}\label{gaugefix}
&&\lim_{\alpha \to 0}\frac{1}{2} \Tr \frac{1}{\nabla^2_{\gamma} \,
\hat\Gamma_{k}[\bar g]+
S_{\rm gf}^{(2)}[\bar{g}] + \mathcal{R}_k^{\rm grav}[\bar g] } \,
\partial_t \mathcal{R}^{\rm grav}_k
[\bar g]\,- \nonumber\\
&&\quad\quad-\Tr \,\frac{1}{\,-\,\mathcal{Q}[\bar g]+
\mathcal{R}_k^{gh}[\bar g] } \,\partial_t \mathcal{R}^{\rm gh}_k[\bar g]\,. 
\end{eqnarray}
\Eq{gaugefix} entails the reduction of the propagator in terms of
covariant derivatives w.r.t.\ Vilkovisky's connection to an
expression which depends on Riemannian covariant derivatives. The
price to pay is the intermediate introduction of a gauge fixing,
which, however, does not play a role in the final
expression.

\section{Approximation}\label{sec:approximation}

The standard Einstein-Hilbert truncation in the background field
approach to quantum gravity amounts to introducing a flowing
cosmological constant and Newton constant into the full, gauge-fixed
Einstein-Hilbert action, \eq{eq:finalS}, that is 
\begin{eqnarray}\label{eq:EHfull}
  &&\hspace{-.7cm}\Gamma_{\tiny \mbox{EH}}[\bar g;\phi] = 
 2 \kappa^2 \bar Z_{N,k}\int d^d x\sqrt{g}\,
  \left(-R(g)+2\bar \Lambda_k\right)+S_{\rm gf}\nonumber\\
  &&\hspace{-.5cm}+2\int d^d x\sqrt{g}\,\bar{C}_{\mu} 
  \Big(\delta^{\mu}_{\nu}\Delta+
  R^{\mu}_{\nu}+(1+\theta)\nabla^{\mu}\nabla_{\nu}\Big) C^{\nu}\,,
\end{eqnarray}
where $\kappa^2$ is defined in \eq{eq:kappa}, and $\hat\Gamma_{\tiny
  \mbox{EH}}=\Gamma_{\tiny \mbox{EH}}+S_{\rm gf}$. As before $\,h = -
\sigma[\bar{g};g]\,$ and all geometric quantities such as $\Delta$,
$R^{\mu}_{\nu}$ and $\nabla^{\mu}$ are constructed with respect to the
full metric $g$. The cut-off dependent quantities $\bar Z_{N,k}$ and
$\bar \Lambda_k$ stand for the scale-dependent wavefunction
renormalisation factor and the scale-dependent cosmological constant
respectively. At vanishing fluctuation field $h=0$ \eq{eq:EHfull}
solely depends on the full metric $g=\bar g$ and is
diffeomorphism-invariant. At $h\neq 0$ it is still
diffeomorphism-invariant w.r.t.\ a combined transformation of $\bar g$
and $h$. However, the approximation \eq{eq:EHfull} does not respect
the Nielsen identity, \cite{Pawlowski:2003sk,Pawlowski:2005xe}. This
is discussed in detail in the next Section~\ref{sec:Nielsen}.

Here we simply anticipate the occurrence of further terms due to the
Nielsen identity and introduce an extended Einstein-Hilbert truncation,
\begin{equation}
  \Gamma_k[\bar g;h,\bar C,C]=\Gamma_{\tiny \mbox{EH}}^{\ }[g;\bar C,C] 
  + \Delta\Gamma[\bar g;h]+ {\rm higher\ order}\,.
\label{eq:EH}
\end{equation}
with 
\begin{equation}
\Delta \Gamma [\bar g;0]\,= \, 0\,, 
\end{equation}
The higher order terms stand for additional diffeomorphism-invariant
terms in the full metric $g$, and the ghost part of the action is the
same as in eq.\eq{eq:EHfull}. The Einstein-Hilbert term depends on the full metric
$g$ whereas $\Delta\Gamma[\bar g;h]$ stands for quantum fluctuations
that depend on the fluctuations $h$ and the background $\bar g$
separately. In the minimally consistent completion of the
Einstein-Hilbert truncation $\Delta\Gamma$ contains a 'mass' term for
the fluctuation field $h$ and a contribution to the kinetic term for
$h$. In DeWitt's condensed notation the decomposition has the form
\begin{equation}\label{eq:h^2}
  \Delta\Gamma[\bar g;h] =\Delta \Gamma_1 \, +
  \, \Delta \Gamma_2=\Delta\Gamma^a h_a+\012 
  \Delta\Gamma^{ab} h_a h_b\,, 
\end{equation}
with symmetric coefficients $\Delta\Gamma_{ab}=\Delta\Gamma_{ba}$. 
The term $\Delta \Gamma_1$ is linear in $\,h\,$ and reads
\begin{equation}
\Delta\Gamma_1[\bar g;h] = \int d^d x \sqrt{g}\,  
\Delta\Gamma^{\mu\nu}[g] h_{\mu\nu}\,. 
\end{equation}
whereas $\Delta \Gamma_2$ is quadratic in $h$ with 
\begin{equation}
\Delta \Gamma_2 =\frac{1}{2} \int d^d \sqrt{g} \, h_{\mu\nu} \Delta 
\Gamma^{\mu\nu\rho\sigma}[g] \, h_{\rho\sigma} \,\,.  
\end{equation}
Here we dropped terms of order higher than $h^2$ with $g$-dependent
expansion coefficients $\Delta \Gamma^{a_1\cdots a_n}$. Due to
diffeomorphism invariance the expansion coefficients can only couple
to the $h_A\,$, the fibre variables $h_\alpha$ have to drop
out. Consequently $\Delta\Gamma^a$ and $\Delta\Gamma^{ab}$ have to be
proportional to $\Pi$. The input in the flow equation is the second $h$-
derivative of $\Delta \Gamma$.  For $h=0$, that is $g=\bar
g$, it reads schematically
\begin{eqnarray}\label{eq:DeltaGamma2}
  \Delta\Gamma^{(2)}[g]= \Delta\Gamma_{a,b}+\Delta\Gamma_{b,a}+\Delta\Gamma_{ab}\,,
\end{eqnarray} 
where the first two terms on the rhs arise from $\Delta\Gamma_1$, and
the last term on the rhs comes from $\Delta\Gamma_2$. The
distribution kernel of the second order term is specified with
\begin{equation}\label{eq:DeltaG}
\Delta\Gamma^{(2)}[g] = 4\kappa^2 (Z_N \Lambda_k- \bar Z_N \bar 
\Lambda_k) T_\Lambda  + 2 \kappa^2(Z_N-\bar Z_N) T_N \,.  
\end{equation} 
The $T_\Lambda$ and $T_N$ stand for the tensor structures arising from
the second variation w.r.t.\ $g$ of the cosmological constant term and
the curvature term in the Einstein-Hilbert action in \eq{eq:EH}. The
term \eq{eq:DeltaG} involves two new flowing coefficients $Z_{N,k}$ and
$\Lambda_k$.  Due to \eq{eq:DeltaGamma2}, $\Delta\Gamma^{(2)}[g]$ has
contributions both from $\Delta \Gamma_1$ and $\Delta \Gamma_2$. With
the tensor structure defined by $T_\Lambda$ and $T_N$, \eq{eq:DeltaG}
projects onto the diffeomorphism-invariant variables $h_A$ as demanded
by diffeomorphism invariance.

Note also that the above approximation includes an Einstein-Hilbert
term $\Gamma_{\tiny \mbox{EH}}[\bar g]=\Gamma_{\tiny \mbox{EH}}[\bar
g;0]$. Such a term can be expanded about the full metric, $\bar
g=g-h+O(h^2)$ and is absorbed in the Einstein Hilbert term as well as
in $\Delta \Gamma$. Schematically the expansion of $\Gamma_{\tiny
  \mbox{EH}}[\bar g]$ reads
\begin{eqnarray}\nonumber 
 \Gamma_{\tiny \mbox{EH}}[g-h+O(h^2)]&=&\Gamma_{\tiny \mbox{EH}}[g]
  +\Gamma_{\tiny \mbox{EH},a}[g] h^a+O(h^2)\\[1ex]\nonumber 
&=&\Gamma_{\tiny \mbox{EH}}[\bar g]
  +\Gamma_{\tiny \mbox{EH},a}[g] h^a+ O(h^2)\,.
\end{eqnarray} 
In the same spirit it was possible to introduce a single metric
dependence in $\sqrt{g} \,\Delta\Gamma^{\mu\nu\rho\sigma}[g]$ in
\eq{eq:h^2}. Differences in $\bar g$ and $g$ are absorbed in terms of order 
higher than two in $h$.  The latter are not taken into account in the present
approximation. 

In summary, the minimally consistent  Einstein-Hilbert approximation
leads to the following identity for the second derivative of the effective
action w.r.t.\ $h$,
\begin{equation}\label{eq:approxprop}
  \Gamma_{k,ab}[\bar g,0]=\left.\left(\hat\Gamma_{\tiny\mbox{EH}}
    \right)_{\cdot cd}^{\ }[\bar g] 
    \,{\Pi^c }_a\, {\Pi^d }_b \right|_{\bar\Lambda\to \Lambda,\bar Z_N\to Z_N}+S^{(2)}_{\rm gf}\,,
\end{equation} 
see \eq{eq:hA}, \eq{eq:EH} and \eq{eq:h^2}. This leaves us with the
task to compute $\nabla_\gamma^2 \hat \Gamma_{\tiny\mbox{EH}}+
S^{(2)}_{\rm gf}$ at $g=\bar g$. The results are listed in
Appendix~\ref{app:hatG2+Sgf2}. The propagator on the
right hand side of the flow equation \eq{eq:flow} depends on
$\Gamma_{k,ab}$ and hence only on $(\Lambda,Z_N)$. The
standard background field approximation in the geometrical approach
amounts to
\begin{equation}\label{eq:backapprox} 
(\Lambda,Z_N)=(\bar\Lambda,\bar Z_N)\,, 
\end{equation} 
in \eq{eq:approxprop}. In other words, the additional term
$\Delta\Gamma$ simply compensates for the fact that the propagator of
the fluctuation field does not depend on the background parameters
$\bar\Z_N,\bar \Lambda$ but on the fluctuation parameters
$\Z_N,\Lambda$. 

We summarise the flow equation in the approximation introduced above
as follows: we have a coupled set of differential equations for the
dimensionless pair $(g_N,\lambda)$ of dynamical couplings with 
\begin{eqnarray}\label{eq:defofdimless}
  g_N= \0{k^{d-2} G_N}{Z_{N,k}}\,,\quad \lambda=k^{-2}\Lambda_k\,,
\quad \eta_N=-\0{\partial_t Z_{N,k}}{Z_{N,k}}\,, 
\end{eqnarray}
leading to 
\begin{eqnarray}\label{eq:etaNg}
  \eta_N = \0{\partial_t g_N +(2-d) g_N}{g_N}. 
\end{eqnarray}
With the definitions in \eq{eq:defofdimless} we have  
\begin{subequations}\label{eq:fullflow}
\begin{eqnarray} \label{eqflowg}
\partial_t g_N +(2-d) g_N&= &F_g(g_N,\lambda) \,,\\[1ex]
\partial_t\lambda  +(2-\eta_N) \lambda &= &F_{\lambda}(g_N,\lambda)\,. 
\label{eq:flowlambdafluc}
 \end{eqnarray} 
\end{subequations}
The set of flow equations \eq{eq:etaNg} does not depend on the
background couplings $(\bar g_N,\bar \lambda)$ defined analogously 
to  \eq{eq:defofdimless}
\begin{eqnarray}\label{eq:backdefofdimless}
  \bar g_N= \0{k^{d-2} \bar G_N}{Z_{N,k}}\,,\quad \bar \lambda=k^{-2}\bar \Lambda_k\,,
\quad \bar \eta_N=-\0{\partial_t \bar Z_{N,k}}{\bar Z_{N,k}}\,.  
\end{eqnarray}
This fact reflects the background independence of the approach.  In turn,
the flows \eq{eq:fullflow} induce flows for the dimensionless pair $(\bar g_N
,\bar \lambda)$
\begin{subequations}\label{eq:fullflowbar}
  \begin{eqnarray} \label{eqflowgbar} \0{g_N}{\bar
      g_N}\left(\partial_t {\bar g}_N +(2-d) \bar g_N\right)
    &= &\bar F_g(g_N,\lambda) \,,\\[1ex]
    \0{g_N}{\bar g_N}\left(\partial_t\bar \lambda +(2-\bar \eta_N)
      \bar \lambda\right) &= &\bar F_{\lambda}(g_N,\lambda)\,.
\label{eq:flowlambdabar} 
\end{eqnarray}
\end{subequations}
Note that the right-hand sides in \eq{eq:fullflowbar} do not depend on
$\bar g_N,\bar\lambda$, and thus the flow of the
background couplings $(\bar Z_N,\bar \Lambda)$ only depends on the
dynamical couplings $(Z_N, \Lambda)$. The ratios $g_N/\bar g_N$ on the
lhs of \eq{eq:fullflowbar} simply originate from using the Newton
constants $g_N, \bar g_N$ instead of $Z_N,\bar Z_N$.

In the background field approximation, \eq{eq:approxprop}, the system
of flows \eq{eq:fullflow},\eq{eq:fullflowbar} is substituted by
\eq{eq:fullflowbar} with $g_N=\bar g_N$ and $\lambda =\bar
\lambda$. In this case the ratio is unity and we arrive at a coupled
set of two flow equations for $g_N$ and $\lambda$ very similar to the
standard background flows, see Section~\ref{sec:backgroundflow}. The only
difference is the appearance of the covariant derivatives in the
propagator. In the present work we shall also solve the full system
\eq{eq:fullflow},\eq{eq:fullflowbar}. This is done in the 
Sections~\ref{sec:geobackresults}, \ref{sec:geoflucresults}. 

We close this section with a discussion of observables. In the
background field approach only the correlation functions of the
background metric are diffeomorphism-covariant and can be directly
used to construct observables such as cross-sections. In the present
approach the corresponding correlation functions depend on to the
running Newton constant $\bar g_N$ and the running cosmological
constant $\bar \lambda_N$. In the geometrical approach also the
dynamical couplings $g_N$ and $\lambda$ are coefficients of
diffeomorphism-invariant terms. However, the background couplings
comprise {\it local} information about the theory whereas the
dynamical couplings do not.  A direct physics interpretation has to be
taken with caution. Note, however, that the fixed points of the
theory are signaled by vanishing $\beta$-functions of the dynamical
couplings.

\section{Nielsen identities}\label{sec:Nielsen}
For the computation of $\Delta\Gamma$ we shall resolve the difference
between the background metric and the full metric in a leading order
approximation. To that end we first discuss the usual background
field approach, where $g=\bar g+h$. This relates to the linear
approximation in the geometrical approach. Note, however, that 
the effective action is not a function of $g$ but of $\bar g$ and $h$
separately. The standard approximation used in background field flows
is done by evaluating the flow \eq{eq:flow} at vanishing  $h=0$.  
Then, the flow is a flow for $\Gamma_k[\bar g; 0]$. It
is not closed as the right hand side of \eq{eq:flow} depends on
$\Gamma^{(2)}_k$, the second derivative of the scale-dependent
effective action w.r.t.\ the fluctuation field $h$. The approximation
\begin{equation}\label{eq:approxbar}
  \0{\delta^2 \Gamma_k}{\delta h^2}[\bar g;0]=
  \0{\delta^2  \Gamma_k}{\delta \bar g^2}[\bar g;0]\,, 
\end{equation} 
closes the flow \eq{eq:flow} in the linear approximation. The identity
\eq{eq:approxbar} is violated by the fact that the effective action is
not a function of $\bar g+h$, but of both fields separately. The
truncation \eq{eq:approxbar} fails already at one loop in the standard
background field approach. Hence a computation of the flow of
$\Gamma_k[\bar g,0]$ with \eq{eq:approxbar} deviates from the full
flow already at two loop
\cite{Pawlowski:2001df,Litim:2002xm,Litim:2002ce,Litim:2002hj}, for
infrared diverging regulators it even fails at one loop
\cite{Litim:2002ce}. In \cite{Pawlowski:2001df} the difference to the
correct one loop result for $\Gamma_k^{(2)}$ was used for deriving
the two loop $\beta$-function in Yang-Mills theory. Using the
fluctuating propagators in the flow is also crucial for deriving
confinement within Landau gauge QCD, see \cite{Braun:2007bx}. Indeed,
generally derivatives w.r.t.\ $g$ (or $\,h$) and that w.r.t.\ $\bar g$
are related by Nielsen identities
\cite{Pawlowski:2003sk,Pawlowski:2005xe}. Within the geometrical
approach used in the present work they read
\begin{equation}\label{eq:Nielsen}
  \Gamma_{k,i}+\Gamma_{k,a}\langle \hat h^a{}_{;i}\rangle= 
  \s012 
  G^{ab}\, \mathcal{R}_{ba,i}+\mathcal{R}_{ab}\, G^{bc}
  \s0{\delta}{\delta \bar h^c} \langle \hat h^a{}_{;i}\rangle\,. 
\end{equation}
The subscript $,i\,$ stands for the usual derivative and $\,;i\,$ for the 
$\Gamma_V$-covariant derivative acting on the background metric $\,\bar g$. 
The index $a$ indicates differentiation with respect to the Gaussian normal 
coordinate $\,h^a$ and $G$ stands for the full propagator. \Eq{eq:Nielsen} 
entails that, up to regulator effects, derivatives w.r.t.\ the background metric 
are indeed proportional to those w.r.t.\ geodesic normal fields $h$ as opposed 
to the corresponding identities in the standard background field approach, see 
\cite{Pawlowski:2001df,Litim:2002ce,Pawlowski:2005xe}.

The proportionality factor $\langle \hat h^a{}_{;i}\rangle$ is sensitive to 
quantum effects and encodes the quantum deformation of diffeomorphism invariance 
in a similar way as the BRST master equation encodes the quantum deformation of 
classical BRST invariance \cite{Pawlowski:2003sk,Pawlowski:2005xe}. Inserting the
Einstein-Hilbert truncation \eq{eq:EH} in the Nielsen identity
\eq{eq:Nielsen} we are led to
\begin{eqnarray}\nonumber 
  && \hspace{-1.1cm} {  \Gamma_{{\tiny \mbox{EH}}}^{\ }}_{,a}
  \Bigl(  \langle \hat h^a{}_{;i}\rangle- h^a{}_{;i}\Bigr)
  +\left(\Delta\Gamma_{k,i}+\Delta\Gamma_{k,a} \langle \hat h^a{}_{;i}\rangle
  \right) \\[1ex]
  & & \ \hspace{1cm}=   \s012  G^{ab}\, \mathcal{R}_{ba,i}+\mathcal{R}_{ab}\, 
  G^{bc}  \s0{\delta}{\delta \bar h^c}\langle \hat h^a{}_{;i}\rangle\,.   
\label{eq:NielsenEH}\end{eqnarray}
In \eq{eq:NielsenEH} we have used that the Einstein-Hilbert action
$\Gamma_{{\tiny \mbox{EH}}}^{\ }$ satisfies the classical Nielsen
identity, that is \eq{eq:Nielsen} with vanishing right hand side and
$\langle \hat h^a{}_{;i}\rangle\to h^a{}_{;i}$. This entails
that \eq{eq:NielsenEH} is valid for the general effective action
within the parameterisation $\Gamma_k= \Gamma_{\rm diff}[g,\bar
C,C]+\Delta\Gamma[\bar g;h,\bar C,C]$ with diffeomorphism-invariant
$\Gamma_{\rm diff}$. The present approximation is the simplest case of
such a splitting. In the full quantum case, the replacement $\langle
\hat h^a{}_{;i}\rangle\to h^a{}_{;i}$ is a mean field
approximation,
\begin{equation}\label{eq:mean}
\left. \langle \hat h^a{}_{;i}\rangle\right|_{\rm mean\ field}= h^a{}_{;i}\,.
\end{equation}
Using this approximation in \eq{eq:NielsenEH}, the first term on the
left hand side vanishes and we arrive at
\begin{equation}\label{eq:NielsenEHprefinal}
\Delta\Gamma_{k,i}+\Delta\Gamma_{k,a} \,h^a{}_{;i} 
=  \s012   G^{ab}\, \mathcal{R}_{ba,i}+\mathcal{R}_{ab}\, G^{bc} 
\s0{\delta}{\delta \bar h^c} \,h^a{}_{;i} \,.  
\end{equation}
Note that implicitly the mean field approximation is behind both, the
Einstein-Hilbert approximation as well as the identity
\eq{eq:approxbar}. The quantum deformation of diffeomorphism
invariance encoded in $\langle \hat h^a{}_{;i}\rangle-
h^a{}_{;i}$ can be taken into account successively by the flow of
$\langle \hat h^a{}_{;i}\rangle$, see \cite{Pawlowski:2003sk}. This
is postponed to future publications. 

\Eq{eq:NielsenEHprefinal} can be used to compute the differences
between the background parameters $(\bar\Lambda,\bar Z_N)$ and the
fluctuation parameters $(\Lambda, Z_N)$ in the given Einstein-Hilbert
approximation \eq{eq:EH}.  We shall do this in an expansion about
vanishing geodesic field $h=0$ as well as in an expansion of the
$\bar g$-dependences of the regulator that induce the right hand side of
\eq{eq:NielsenEHprefinal}. At $h=0$ we have
\begin{eqnarray}\nonumber 
&h^c{}_{;a}=-\delta^c{}_a\,,\quad 
h^d{}_{;ca}=0\,, &\\[1ex] 
&h^d{}_{;c(ab)}=\016
(R_V^d{}_{acb}+R_V^d{}_{bca} +2 R_V^d{}_{c(ab)})\,,&
\label{eq:vilexp}\end{eqnarray}
the third derivative of $h$ is proportional to the affine part of the
curvature tensor $R_V$ of Vilkovisky's connection. The $R_V$-terms
would contribute to a further $h$-derivative of $\Delta\Gamma_{a}[\bar
g]$. In the present work we drop them as sub-leading. Then, the last
term in \eq{eq:NielsenEHprefinal} vanishes and we conclude that
\begin{equation}\label{eq:Gamma1}
\Delta\Gamma_a[\bar g]  
=- \s012 G^{cd}\, \mathcal{R}_{dc,a}[\bar g]\,.
\end{equation} 
This fixes the first two terms on the rhs of \eq{eq:DeltaGamma2} which
are computed with a further (covariant) derivative of \e the first
term on the rhs derives from $\Delta\Gamma_2$, \eq{eq:Gamma1}
w.r.t. $\sqrt{g}$. Now we take a $h^b$-derivative of
\eq{eq:NielsenEHprefinal} at fixed $g$. Evaluated at
$h=0$ this reads
\begin{equation}\label{eq:Gamma}
  \Delta\Gamma_{ab}[\bar g]=-\Delta\Gamma_{b,a}[\bar g]-\012 \0{\delta}{\delta h^b}
  \left( G^{cd}\, \mathcal{R}_{dc,a}\right)[\bar g]\,,
\end{equation}
where the $h^b$-derivative of the flow term at fixed $g$ only hits the
cut-off terms and the $\bar g$-dependences of $\Delta\Gamma^{(2)}$ in
the propagator $G$, evaluated at $h=0$. In combination this leads to a
$\bar g$-derivative of $-\Delta\Gamma_{a}[\bar g]$ at fixed
regulator. In terms of the regulator-induced $\bar g$-dependences this
is the leading term in the Nielsen-identity. Hence, within leading
order we arrive at
\begin{equation}\label{eq:Gamma+}
  \Delta\Gamma_{ab}[\bar g]= -\Delta\Gamma_{(b,a)}[\bar g] 
  \equiv - \frac{1}{2}\,\Delta\Gamma_{b,a}[\bar g]
  -\frac{1}{2}\,\Delta\Gamma_{a,b}[\bar g]\,.
\end{equation}
The results \eq{eq:Gamma+} and \eq{eq:Gamma1} allow us 
to compute $\Delta\Gamma^{(2)}$ defined in \eq{eq:DeltaGamma2}, 
\begin{eqnarray}
  \Delta\Gamma^{(2)}_{ab}& =& -\s012 (\,  G^{cd}\, 
  \mathcal{R}_{dc,(a}\,)_{,b)}\label{eq:DeltaGamma2ex}\\[1ex]
  & =& -\s012 (\,  G^{cd}\, \mathcal{R}_{dc}\,)_{,ab}+
  \s012 (  G^{cd}{}_{,(a}\, \mathcal{R}_{dc})_{,b)}\,, \nonumber
\end{eqnarray}
where as usual the parenthesis indicate symmetrization.
For the flow of the propagator of the dynamical fields $h$ on the
right hand side of \eq{eq:Gamma2} we sum-up the $t$-derivative of
\eq{eq:Gamma2} and $\partial_t \Gamma_k[g]_{,ab}$. The latter
expression is the second derivative w.r.t. $\bar g$ of the flow
\eq{eq:flow}, evaluated at $h=0$. Finally we need the covariant
derivatives with the Vilkovisky connection of $\Gamma_{\tiny \mbox{EH}}$
at vanishing fluctuation field $h=0$. For the second derivative of
$\Delta\Gamma$ at $h=0$ this does not make a difference and we arrive
at
\begin{equation}
\left. \partial_t\Gamma_{k;ab}\right|_{h=0} = - \s012 ( \mathcal{R}_{dc} 
\partial_t G^{cd}\, )_{;ab}+
\s012 \partial_t (  \mathcal{R}_{dc}\, G^{cd}{}_{(;a})_{;b)}\,. 
\label{eq:dtGamma2}
\end{equation}
The first term on the right hand side is a total second derivative
w.r.t.\ $\bar g$, and can be computed with heat kernel techniques
analogously to the standard flow. In turn, the second term is not that
easily accessible. However, it can be minimised by an appropriate
regulator choice and will be discussed in the next Section.  In
summary the minimal consistent Einstein-Hilbert truncation \eq{eq:EH}
together with the flow of the fluctuation two-point function
\eq{eq:dtGamma2} allows us to compute the flow of all parameters,
$(g_N,\lambda)$ and $(\bar g_N,\bar \lambda)$, in the given
approximation.

\section{Phase diagram in the standard background field
  approximation}\label{sec:backgroundflow}

\begin{figure}[t]
  \includegraphics[width=.88\columnwidth]{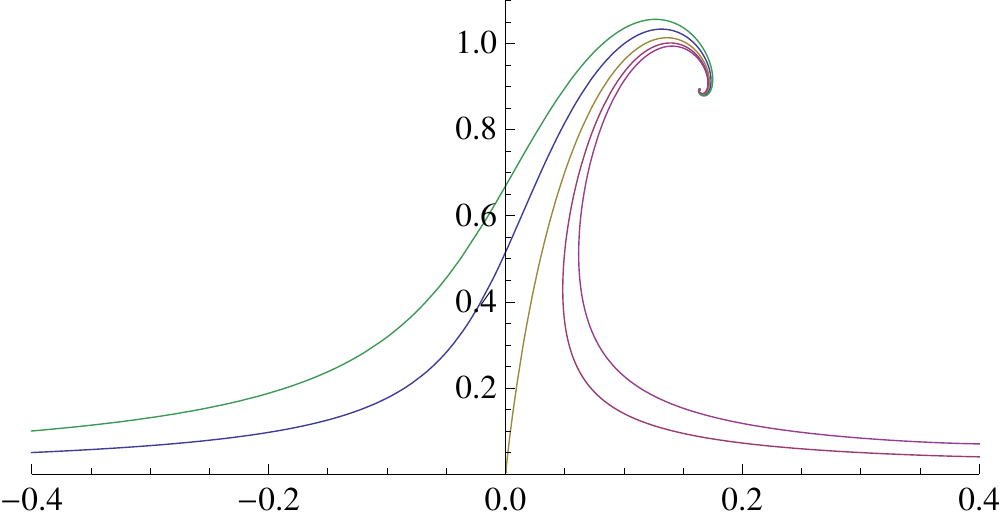}
\put(-130,105){\large $g_k$}
\put(2,5){\large $\lambda_k$}
  \caption{Phase diagram in the background field approximation,
    \eq{eq:backg}, \eq{eq:backlambda}, with the UV fixed point
    $({g_{N}}_*,\lambda_{*}) =(0.893,0.164)$.}
\label{fig:back}
\end{figure}

We are now in the position to compute the flow of the couplings in the
extended Einstein-Hilbert approximation put forward in
Section~\ref{sec:approximation}. Given the close relation between the
geometrical effective action and the background effective action in
the Landau-DeWitt gauge, it is also worth discussing the similarities
and the differences to the latter, see e.g.\ the reviews
\cite{Reuter:2007rv,Percacci:2007sz,Litim:2011cp,Reuter:2012id} and the literature
therein. Hence, for illustrative purposes and for the sake of
comparison with results in the literature we first solve the
geometrical flow equation in the background field approximation. This
also allows us to disentangle the effects of this approximation
from those arising from symmetry constraints.  The
background field approximation is implemented by identifying $g_N
\equiv \bar{g}_N$ and $\lambda \equiv \bar{\lambda}$, see
\cite{donkin}. This is achieved by taking $\,\Delta \Gamma[\bar{g},h]
= 0\,$ in eq.(\ref{eq:h^2}) and evaluating the Einstein-Hilbert action
$\Gamma_{\tiny \mbox{EH}}[g;C,\bar{C}]$ at the background geometry
$g=\bar{g}$ and vanishing ghosts. Then the flow equation reduces to
\begin{eqnarray}\label{eq:flowdipl}
  \partial_t \Gamma_{\tiny \mbox{EH},k}[\bar g]&=&-\Tr \,\frac{1}{\,-\,\mathcal{Q}[\bar g]+
\mathcal{R}_k^{gh}[\bar g] } \,\partial_t \mathcal{R}^{\rm gh}_k[\bar g]\\ \no 
&& \hspace{-1.5cm} +\frac{1}{2} \Tr \frac{1}{\,\nabla^2_{\gamma} \,
    \Gamma_{\tiny \mbox{EH}, k}[\bar g]+
    S_{\rm gf}^{(2)}[\bar{g}] + \mathcal{R}_k^{\rm grav}[\bar g] } 
  \,\partial_t \mathcal{R}^{\rm grav}_k
  [\bar g] \,,
\end{eqnarray}
where $\mathcal{Q}$ is the ghost operator and the limit $\alpha\to 0$
is implied. The traces in \eq{eq:flowdipl} only sum over momenta and internal indices. 
Following the common philosophy within the background
field approach, we make the extra assumption that the gauge-fixing
term has a $Z$-dependence, i.e:
\begin{equation}\label{eq:Zalphaback}
S_{\rm gf} = \frac{\bar{Z}_{N,k}\kappa^2}{4\alpha}\,\int d^d x 
\sqrt{\bar{g}} \, \bar{g}^{\tau\lambda}\,\mathcal{F}^{\rho'\sigma'}_{\tau}[\bar{g}]\,
\hat h_{\rho'\sigma'}\,\mathcal{F}^{\mu\nu}_{\lambda}[\bar{g}]\,\hat h_{\mu\nu} \,.
\end{equation} 
This additional approximation \eq{eq:Zalphaback} is resolved in the
next Section~\ref{sec:dynflow}.  The rest of the present calculation
proceeds by employing the standard York transverse-traceless
decomposition detailed in Appendix~\ref{app:york} and choosing regulators
whose tensor structure is adapted to this decomposition.  The full set
of regulators is listed in Appendix D. Here we explicitly provide the results 
for the optimised shape function, \cite{Litim:2000ci},  
\begin{equation}\label{eq:optimised}
r(x)=(1-x)\,\theta(1-x)\,. 
\end{equation} 
With \eq{eq:optimised} the computations are much simplified. We use
the York decomposition detailed in the
Appendices~\ref{app:york}, \ref{app:hatG2+Sgf2} and the corresponding
regulators in Appendix~\ref{app:regulators} with the optimised shape
function \eq{eq:optimised}. The relevant threshold functions for the
optimised regulator are evaluated in Appendix~\ref{app:thresopt}. The
resulting flow equations read in four dimensions, d=4,
\begin{eqnarray}
&& \hspace{-.3cm} \partial_t g_N -2 g_N=\\  \nonumber 
&&\hspace{-.2cm} -\0{g_N^2}{\pi}\frac{\frac{5}{3}+\frac{2}{3}\,(1-2\lambda)+
\frac{25}{24}(1-2\lambda)^2}{(1-2\lambda)^2 -\frac{g_N}{2\pi}
\left[\frac{5}{9}+\frac{1}{3}(1-2\lambda)-\frac{5}{12}(1-2\lambda)^2\right]}  \,,
 \label{eq:backg}\end{eqnarray} 
and 
\begin{eqnarray}
&& \hspace{-.3cm} \partial_t \lambda +2 \lambda =\\\nonumber 
&&\hspace{-.2cm} \eta_N \left(\lambda -
\frac{g_N}{4\pi}\left(\frac{2}{3}+\frac{1}{1-2\lambda}\right)\right) 
-\frac{g_N}{4\pi}
\left(4-\frac{6}{1-2\lambda}\right)\,,  \label{eq:backlambda}
\end{eqnarray}
with $\eta_N$ defined in \eq{eq:defofdimless}, \eq{eq:etaNg}. The two
flow equations \eq{eq:backg} and \eq{eq:backlambda} define $\bar F_g$
and $\bar F_\lambda$ in \eq{eqflowgbar} and \eq{eq:flowlambdabar}
respectively in the standard background field approximation. The respective
flow diagram is given in Fig.~\ref{fig:back}.
The flow equations \eq{eq:backg},\eq{eq:backlambda} admit an
attractive UV fixed point at
\begin{equation}\label{eq:FPback}
({g_{N}}_*,\lambda_{*}) =(0.893,0.164)\,, 
\end{equation}
and a repulsive Gaussian fixed point at the origin,
$({g_{N}}_*,\,\lambda_{*}) =(0,0)$, see also Fig.~\ref{fig:back}.  
We emphasize that \eq{eq:backg} and \eq{eq:backlambda} are completely
independent of the gauge-fixing parameter $\alpha$. This is to be
expected as the geometrical flow itself, by definition, does not
depend on the gauge-fixing condition. This is in clear
contradistinction to $\alpha$-dependence observed in the usual
background field approach, see \cite{Lauscher:2001ya,Litim:2003vp} and
the reviews \cite{Reuter:2007rv,Percacci:2007sz,Litim:2011cp,Reuter:2012id}.

Note also that the geometrical effective action at vanishing
fluctuation field $h=0$ can be linked to the background effective
action in the Landau-DeWitt gauge. In the present approach this matter
is complicated due to the presence of the regulator and the
approximation involved. We observe that in four dimensions, $d=4$, the flow
equations \eq{eq:backg} and \eq{eq:backlambda} indeed agree with the
background flows with the optimised regulator \cite{Litim:2003vp}; however in
dimensions $d\neq 4$ the flows do not agree. The parameter $a_2$ reads
in the background field approach
\begin{equation} \label{eq:backa2}
a_2= -\0{d^3-4d^2+7 d-8}{2 (d-1)}\,, 
\end{equation}
and differs from $a_2$ in the geometrical approach, see
\eq{eq:a12}. This is not unexpected as the fluctuation fields in the
present approach are non-polynomially related to the linear
fluctuation fields in the standard background field approach.

\section{Dynamical Flows}\label{sec:dynflow}

In this section we set-up the full dynamical flow for the fluctuation
field coefficients $g_N$ and $\lambda$. The results apply to general
background flows including the standard background approach. The
latter can be obtained within the approximation discussed in the
previous section, as well as in approximations going beyond 
\eq{eq:Zalphaback}.  For the dynamical flow we have to evaluate
\eq{eq:dtGamma2}. This can be done for general regulators with
off-shell heat kernel techniques, see e.g.\
\cite{Benedetti:2010nr}. Here, however, we shall employ a specific
choice of the regulator which removes the second term on the rhs,
and also further facilitates the computations. To that end we choose
the (partially) optimised regulator
\cite{Litim:2000ci,Pawlowski:2005xe} in \eq{eq:regulators} with the
shape function in \eq{eq:optimised}. This regulator renders all
propagators constant for spectral values below the cut-off
scale. Above the cut-off scale the regulator vanishes and hence ${\cal
  R }\partial_t G_{;a} \equiv 0$. Then \eq{eq:dtGamma2} reduces to a
total (covariant) derivative,
\begin{equation}
\left. \partial_t\Gamma_{k;ab}\right|_{h=0} = - \s012 ( \mathcal{R}_{dc} 
\partial_t G^{cd}\, )_{;ab}\,. 
\label{eq:dtGamma2ex}
\end{equation}
\Eq{eq:dtGamma2ex} can be computed with standard heat kernel techniques. Note
that the optimisation in \eq{eq:regulators} serves a
twofold purpose. First, it is the regulator choice partially adapted
to the approximation specified in Section~\ref{sec:approximation} and
hence maximises the physical content of the approximation. It only is
a partial optimisation as the present choice does not fully resolve
the issue of relative cut-off scales discussed in
\cite{Pawlowski:2005xe}. This entails that only the choice $\theta
\approx -1$ is optimised. Second, the choice \eq{eq:optimised} allows
us to relate fluctuation flows and background flows with simple
algebraic identities as will be shown below.

As a showcase for the full computation we shall evaluate the flow for
$\Lambda_k$ on a flat background $\bar g=\eta$. This amounts to
evaluating \eq{eq:dtGamma2ex} on that background. However, as the rhs is
explicitly a total second derivative it can be easily integrated. This leads us to 
\begin{equation}
 \partial_t\left( Z_{N,k} \,\Lambda_k \right) =-\01{{4 \kappa^2 \rm Vol}} 
  \Tr \,\mathcal{R}_k[\eta] \,\partial_t
  \0{1}{\Gamma_k^{(2)}[\eta;0]+ \mathcal{R}_k[\eta] } \,,  
\label{eq:difflambda1}
\end{equation}
where ${\rm Vol} $ stands for the volume factor $\int d^4 x \sqrt{\bar
  g}$.  It also occurs in the trace on the right hand side and hence
drops out. With \eq{eq:difflambda1} we also get a simple relation
between the flow of $ Z_{N,k} \,\Lambda_k$ and that of $\bar Z_{N,k}
\,\bar \Lambda_k$,
\begin{eqnarray}\label{eq:lambdalambdabar}
 \partial_t\left( \bar Z_{N,k} \,\bar \Lambda_k \right)&=&  
\partial_t\left( Z_{N,k} \,\Lambda_k \right)\\\nonumber  
&&\hspace{-.8cm}+ \01{{4 \kappa^2 \rm Vol}} 
\partial_t \left( \Tr \,\mathcal{R}_k[\eta] \, \0{1}{
    \Gamma_k^{(2)}[\eta;0]+ \mathcal{R}_k[\eta] }\right) \,. 
\end{eqnarray}
For the regulator \eq{eq:optimised} the integrands in the traces in
\eq{eq:difflambda1} and \eq{eq:lambdalambdabar} agree up to
prefactors. For the integrands in \eq{eq:difflambda1} we have a simple
relation. For later purpose we already write it in its general form in
the presence of a non-vanishing curvature,
\begin{eqnarray}\nonumber 
  &&  \hspace{-1.4cm}\mathcal{R}^{\rm mode}_k \partial_t \left(\0{1}{k^2 Z} \0{1}{1+b_{\rm mode} 
\rho+c_{\rm mode} \lambda}\right)\\\nonumber 
  &=& -\left(2 -\eta^{\ }_Z +\0{ c_{\rm mode}\dot \lambda-2 b_{\rm mode}\rho}{1
   +b_{\rm mode} 
\rho   +c_{\rm mode} \lambda}\right)\\
&&\times    \mathcal{R}^{\rm mode}_k   
  \0{1}{k^2 Z}\0{1}{1+b_{\rm mode} 
\rho+c_{\rm mode} \lambda}  \,, 
 \label{eq:genform}\end{eqnarray} 
 The quadratic part of the action is detailed in the
 Appendices~\ref{app:hatG2+Sgf2},\ref{app:ghostG2} and the regulator
 ${\cal R}_k^{\rm mode}$ of a given mode is a function of
 $\Delta_{\bar g}$, see Appendix~\ref{app:regulators}. This leads to
 the denominator in \eq{eq:genform}. We also have used the notation
 $\dot \lambda=\partial_t \lambda$, and have introduced the
 dimensionless curvature $\rho$
\begin{equation}\label{eq:rho}
\rho=R/k^2\,, 
\end{equation}
with $\partial_t \rho=-2 \rho$.  In \eq{eq:genform} the coefficient
$c_{\rm mode} = 0,\,- d/(d-2)$ takes into account the
$\lambda$-dependence of the different modes. In turn, the coefficients
$b_{\rm mode}$ are more complicated. However, their specific value
drops out for the computations done here. The different wave function
renormalisations lead to anomalous dimensions
\begin{equation}\label{eq:betas}
  \eta_Z^{\ } =-\partial_t \ln Z\,, \quad \quad \dot \lambda = 
  \partial_t \lambda\,. 
\end{equation} 
Evidently the rhs of \eq{eq:genform} is that of the integrand of the
trace in \eq{eq:lambdalambdabar} up to the prefactor in
parenthesis. In \eq{eq:genform} we have used the fact that the
propagators are flat for $\Delta_{\bar g} < k^2$. The anomalous
dimension $\eta_Z$ is vanishing for ghosts in the present
approximation due to $Z_{\rm gh}\equiv 1$. Moreover, we have
$\eta_Z=\eta_N$ with $Z=Z_N$ for the transversal graviton
modes. For the gauge mode we have $Z_\alpha=1$. It is only
  introduced for convenience and it flow vanishes due to
  diffeomorphism invariance. Then, with \eq{eq:genform} we compute
\eq{eq:difflambda1} as
\begin{eqnarray} \label{eq:lambdaI}
\left( \partial_t +(2-\eta_N)\right)\lambda  &=& 2 ( I_{\lambda,0} +
I_{\lambda,{\rm gh}}) \\[1ex]
& & +\left(2 - \eta_{N}  -\0{\frac{d}{d-2}\dot \lambda}{1-\frac{d}{d-2} 
\lambda} \right)I_{\lambda,-2}\,, \nonumber 
\end{eqnarray}
with 
\begin{equation}\label{eq:Ilambdas}
  I_{\lambda,c_{\rm mode}}=\0{8\pi\, g_N}{k^d \rm Vol}  
  \Tr\,\mathcal{R}^{\rm mode}_k(\Delta_{\bar g})  
  \0{1}{k^2 Z}\0{1}{1+c_{\rm mode} \lambda}\,,
\end{equation}
and similarly for $I_{\lambda,{\rm gh}} $. The term $2
I_{\lambda,0}$ in \eq{eq:Ilambdas} only comes from the gauge mode. The
$I$'s defined in \eq{eq:Ilambdas} also allow us to compute the second
line in \eq{eq:lambdalambdabar}. This term depends on covariant
momenta, the cut-off scale $k$, the normalised cosmological constant
$\lambda=\Lambda_k/k^2$, and its canonical dimension is $d$. Therefore
the trace in the second line in \eq{eq:lambdalambdabar} leads to an
explicit factor $k^d$ multiplied by a function of $\lambda$. The
$t$-derivative reproduces the term as well as a
$\dot\lambda\partial_\lambda$-term. Hence we conclude that
\begin{eqnarray}\label{eq:genform2} 
&&\hspace{-1.0cm} \0{8\pi g_N}{k^d \rm Vol} 
\partial_t \left( \Tr \,\mathcal{R}_k[\eta] \0{1}{
    \Gamma_k^{(2)}[\eta;0]+ \mathcal{R}_k[\eta] }\right) \\[1ex] 
&=& \hspace{.2cm} d \left(I_{\lambda,0}+
  I_{\lambda,\rm gh}\right)+\left(d +\0{\frac{d}{d-2}\dot \lambda}{1
    -\frac{d}{d-2}\lambda}
\right)  I_{\lambda,-2}\,. \nonumber
\end{eqnarray}
Adding \eq{eq:genform2} to \eq{eq:lambdaI} gives the rhs of
\eq{eq:lambdalambdabar}. Resolving this for the flow of $\bar\lambda$
yields
\begin{eqnarray}\nonumber 
 \0{g_N}{\bar g_N}\left( \partial_t +(2-\bar\eta_N)\right)\bar\lambda &=& 
(d+2) \left(I_{\lambda,0} + I_{\lambda,\rm gh}\right)\\[1ex] 
&+&  \left( d+2 -\eta_N\right) \, I_{\lambda,-2}\,. 
\label{eq:dotlambdaI}\end{eqnarray} 
\Eq{eq:dotlambdaI} is the standard background flow if we apply
$(g_N,\lambda)\to (\bar g_N,\bar\lambda)$ on the right hand
side. This allows us to determine the coefficient functions
$I_\lambda$ from the standard background field approximation to the
flow. The coefficient functions $I_\lambda$ are then inserted in the 
flow of $\kappa^2 \Lambda_k$ in \eq{eq:lambdaI}.

Note that even though \eq{eq:dotlambdaI} was derived in a flat
background we have only used general properties and relations for
the flow, and hence \eq{eq:lambdaI},\eq{eq:dotlambdaI} are valid for
arbitrary backgrounds. Indeed we can even extend \eq{eq:dotlambdaI}
to the full flow of the effective action in the Einstein-Hilbert
approximation \eq{eq:EH}, \eq{eq:h^2} for general backgrounds with
constant curvature $R$.  In order to access the curvature term we
first have to discuss \eq{eq:genform} if we want to take derivatives w.r.t. 
$R$. 

For the curvature term \eq{eq:genform2} has dimension $d-2$ and hence
we have $d\to d-2$. This leads to 
\begin{eqnarray}\label{eq:genform2rho} 
&\di\hspace{-.8cm}-\0{g_N}{\rho} \0{8\pi g_N}{k^d \rm Vol} 
\partial_t \left( \Tr \,\mathcal{R}_k[\eta]  \0{1}{
    \Gamma_k^{(2)}[\eta;0]+ \mathcal{R}_k[\eta] }\right)_\rho =&\\[1ex] \nonumber
&\di(d-2) \left(I_{N,0}+
  I_{N,\rm gh}\right)+\left(d-2 +\dot \lambda\partial_\lambda\right)  I_{N,-2}\,,  &
\end{eqnarray}
where the subscript $\rho$ in the first line stands for the projection on the term
linear in the curvature $R$, and $\rho$ is the dimensionless
curvature, see \eq{eq:rho}. Note that \eq{eq:genform2rho} is
insensitive to the explicit occurence of $\rho$ in the propagator. Combining
\eq{eq:genform2rho} with \eq{eq:genform} we see, that the dimensional
counting for the term linear in $\rho$ gives a factor $d-2$ for the
modes without explicit curvature-dependence, $b_{\rm mode}=0$, and a
factor $d$ as for the cosmological constant for the terms with
$b_{\rm mode}\neq 0$. The dimensional prefactor does not depend on
$b_{\rm mode}$, whereas the coefficient $I$ does. Hence we finally arrive at
\begin{eqnarray}\nonumber 
\0{8\pi\, g_N}{k^d \rm Vol} \partial_t 
\Gamma_k[\bar g;0]&=&(d+2) \CI_{\lambda} -\eta_N I_{\lambda,-2}\\
&& \hspace{-1.5cm}-\left(d\,  \CI_{N} -\eta_N\, 
  I_{N,-2}\right) \0{\rho}{g_N}-2\, \CI_{N,1} \0{\rho}{g_N}\,, 
\label{eq:flowbar}\end{eqnarray} 
with 
\begin{eqnarray}\nonumber 
\CI_\lambda &=&I_{\lambda,0} +  I_{\lambda,\rm gh} + I_{\lambda,-2}\,,\\[1ex] 
\nonumber 
\CI_N &=&I_{N,0,0} +  I_{N,\rm gh,0}+ I_{\lambda,-2,0}  \\[1ex]
& & + I_{N,0,1}+ I_{N,\rm gh,1} + I_{\lambda,-2,1} \,.
\label{eq:Is}\end{eqnarray} 
In \eq{eq:Is} the last subscript for the coefficients $I_{N}$
with the values $0,1$ labels vanishing and non-vanishing $b_{\rm
  mode}$, and $\CI_{N,0},\CI_{N,1}$ stand for the respective terms. 
\Eq{eq:flowbar} allows us to read-off the coefficients $I_N$
and $I_\lambda$ from the corresponding background field flows of $\bar
g_N$ and $\bar \lambda$ respectively. With these coefficients we can
derive the flow of $g_N$ and $\lambda$ similarly to
\eq{eq:lambdaI}. These flows can be summarised conveniently in
\begin{eqnarray}\label{eq:flowfluc}
  \hspace{-.2cm}\left.\0{8\pi\, g_N}{k^d \rm Vol} \partial_t 
    \Gamma_{\tiny \mbox{EH}}[g;0]\right|_{\lambda,g_N }
  &=&  2\, \CI  -2 \CI_{N,1} \0{\rho}{g_N}
  \\[1ex]\nonumber 
  && - \left(\eta_N+\dot \lambda\partial_\lambda \right) I_{-2}\,, 
  \nonumber \end{eqnarray} 
where $I_{-2}=I_{\lambda,-2}-I_{N,-2}\rho/g_N$ and $\CI_{N,1}$ stands for the second line in \eq{eq:Is}. 
The total coefficient $\CI$ is given by  
\begin{equation}\label{eq:Ifuns}
\CI(\lambda,g_N)= \CI_\lambda(\lambda,g_N)- \CI_N(\lambda,g_N) \0{\rho}{g_N} \,.
\end{equation}
The lhs of \eq{eq:flowfluc} is the flow of
\eq{eq:EHfull} with $(\bar Z_N,\bar\Lambda)\to (Z_N,\Lambda)$, and the
rhs is projected on the respective terms proportional to $r^0$ and
$r^1$. \Eq{eq:flowbar} and \eq{eq:flowfluc} allow us to compute the
flow of $g_N$ and $\lambda$ from a given background flow computed in
the approximation \eq{eq:approxbar}, where $\bar\lambda\to \lambda$ on
the right hand side of the background flow: the coefficient functions
$I$ are determined from \eq{eq:flowbar} with $\bar Z_N\to Z_N$,
$\bar\Lambda\to \Lambda$ on the left-hand side and then identifying
the terms proportional to $\eta_N$, $\partial_t \lambda$ and the rest
in the flows for $g_N$ and $\lambda$. The coefficient functions $I$
are then used in \eq{eq:flowfluc} which gives us the flow equations
for the fluctuation parameters $g_N,\lambda$ in \eq{eq:fullflow}. The
physical observables $\bar g_N, \bar \lambda$ derive from
\eq{eq:flowbar} with the results for the fluctuation parameters
$g_N,\lambda$ inserted on the right-hand side. This gives us the flow
equations \eq{eq:fullflowbar}.

\Eq{eq:flowfluc} and the above relations complete our truncation: we
use the flow of the effective action at vanishing fluctuation fields
$h=0$ within the standard approximation $\Gamma_{k,a}
h^a{}_{;i}+\Gamma_{k,i}=0$. Furthermore we account for the full
Nielsen identity with \eq{eq:flowfluc} and \eq{eq:flowbar} being
sensitive of the background field dependence in the regulator term. If
we apply \eq{eq:flowfluc} and \eq{eq:flowbar} for general regulators
one has to bear in mind that this implies neglecting those terms in
the Nielsen that are proportional to derivatives of the regulator
$\mathcal{R}_k(x)$ w.r.t the covariant momentum $x$. As has been
argued, they are sub-leading, and indeed they can be minimised by
using regulators that are sufficiently flat.  In summary the
geometrical approach provides us with a fully diffeomorphism-invariant
flow for quantum gravity where we have also good qualitative control
over the difference between fluctuation fields and background
metric. The latter distinction is particularly important for the
background independence of the results.

\section{Phase diagram in the geometrical background field
  approximation} \label{sec:geobackresults}

The present diffeomorphism-invariant setting leads to a further
simplification of the results in the previous
Section~\ref{sec:dynflow}. Due to the projection on transversal metric
fluctuations the coefficient $I_0$ vanishes identically, $I_0\equiv
0$. With standard heat-kernel techniques and the York
transverse-traceless decomposition we arrive after some algebra at the
flows for the background Newton constant,
\begin{eqnarray}
\0{g_N}{\bar g_N}\left( \partial_t +(2-d)\right)\bar g_N = 
\bar F^{(1)}_g-\eta_N \bar F_g^{(2)}\,, \label{eq:flowbackZN}
\end{eqnarray} 
and the background cosmological constant 
\begin{equation}\label{eq:flowbackLa}
\0{g_N}{\bar g_N}  \left( \partial_t +(2-\bar\eta_N)\right)\bar
  \lambda = \bar F^{(1)}_\lambda -\eta_N \bar F_\lambda^{(2)}\,, 
\end{equation} 
The $\bar F^{(1)}g(\lambda,g)$ and $\bar F^{(2)}g(\lambda,g)$
originate in terms in the flows proportional to $\partial_t \,k^2r$
and $\partial_t Z_N$ respectively, and only depend on the dynamical
couplings $\lambda,g$. They are given in terms of the coefficient
functions $I_\lambda$ and $I_N$ which are detailed in
Appendix~\ref{app:Is}. 

Note that we have chosen canonical dimensional factors $d$ in the flow
of the Newton constant $\bar g_N$ in \eq{eq:flowbackZN}. Within the
split in curvature-dependent and curvature-independent modes this
implies
\begin{eqnarray}\nonumber 
\bar F_g^{(1)}&=&d \,\CI_{N}+2\CI_{N,1}\,,\\ \nonumber 
\bar F_g^{(2)}&=&I_{N,-2}\,,\\ \nonumber 
F^{(1)}_\lambda& =& (d+2)\CI_{\lambda} \,, \\
F^{(2)}_\lambda& =& I_{\lambda,-2}\,,
\label{eq:Isplit}\end{eqnarray}
with $\CI_N,\CI_\lambda$ defined in \eq{eq:Is} and the related
coefficient functions $F^{(1)}$, $F^{(2)}$ are given in
Appendix~\ref{app:thresopt}, \eq{eq:Ilaopt},\eq{eq:IZNopt}. The
relations \eq{eq:Isplit} are derived within the optimised regulator,
\eq{eq:optimised}, and the $I$'s in Appendix~\ref{app:thresopt} satisfy 
the relations implied in \eq{eq:Isplit}. The system of flow equations
\eq{eq:flowbackZN}, \eq{eq:flowbackLa} constitutes one of the main
results of the present work. It is a fully diffeomorphism-invariant
flow for the pair of background couplings $(\bar g_N,\bar \lambda)$
beyond the background field approximation.  There is no dependence on
the gauge fixing parameter $\alpha$ and the rhs in
\eq{eq:flowbackZN},\eq{eq:flowbackLa} only depend on the dynamical
couplings $(g_N, \lambda)$. Hence, the solution of
\eq{eq:flowbackZN},\eq{eq:flowbackLa} requires that one solves the
flows of the dynamical couplings first. Moreover, vanishing
$\beta$-functions $\partial_t\bar g_N=0=\partial_t \bar\lambda$
determine fixed point pairs $(\hat g_N, \hat \lambda)$. Note, however,
that the $\beta$-functions of the background couplings signal a fixed
point only if the pair $(\hat g_N, \hat \lambda)$ is a fixed point of
the dynamical flow.

\begin{figure}[t]
   \includegraphics[width=.88\columnwidth]{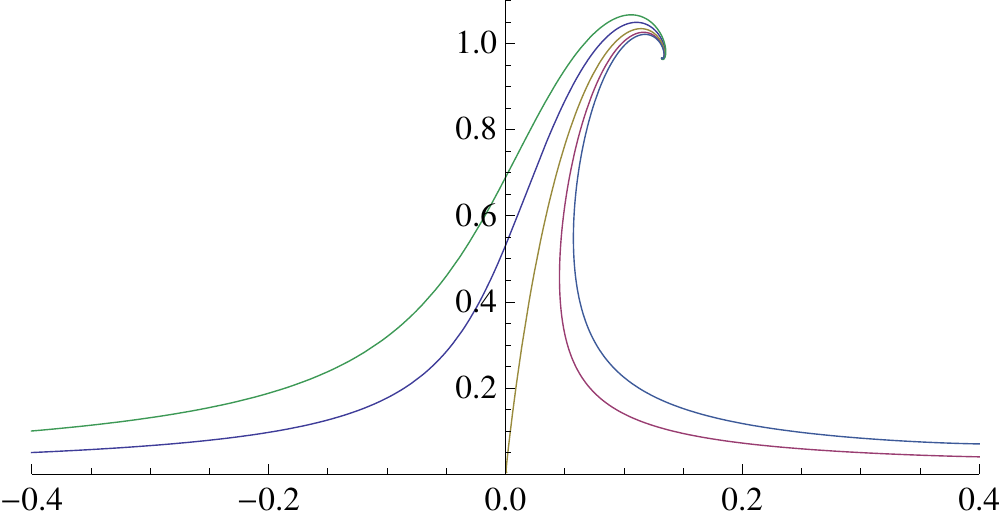}
\put(-130,105){\large $g_k$}
\put(2,5){\large $\lambda_k$}  
 \caption{Improved phase diagram for the background couplings,
     eqs.(\ref{eq:flowbackZNopt}) and (\ref{eq:flowbackLaopt}), with
     an UV fixed point $({g_N}_*,\lambda_{*})=(0.966,0.132)$.}
\label{fig:fullback}\end{figure}

Before we discuss the respective dynamical flows we first implement
once more the background field approximation $(g_N,\lambda)=(\bar
g_N,\bar \lambda)$. Again we use the York decomposition detailed in
the Appendices~\ref{app:york}, \ref{app:hatG2+Sgf2} and the
corresponding regulators from Appendix~\ref{app:regulators} with the
optimised shape function \eq{eq:optimised}. The coefficient functions
$\,I_{\lambda,N}\,$ (with $d=4$) as well as the right hand sides of
eqs.(\ref{eqflowg}) and (\ref{eqflowgbar}) can also be found in
Appendix~\ref{app:thresopt}. With the results of
Appendix~\ref{app:thresopt} we are finally led to the flow equations
for the Newton constant and the cosmological constant,
\begin{equation}
\partial_t g_N-2 g_N= -\frac{g_N^2}{\pi}  \frac{\frac{5}{3}
+\frac{2}{3}\,(1-2\lambda)+
\frac{25}{24}\,(1-2\lambda)^2}{(1-2\lambda)^2 - \frac{g_N}{2\pi}
\left(\frac{5}{9}+\frac{1}{3}\,(1-2\lambda)\right)} \,,\label{eq:flowbackZNopt}  
\end{equation}
and 
\begin{equation}
\partial_t \lambda+2 \lambda =\eta_N \left(\lambda -
\frac{g_N}{4\pi}\frac{1}{1-2\lambda} \,\right)-\frac{g_N}{4\pi}\left(4
-\frac{6}{1-2\lambda}\right)\,, \label{eq:flowbackLaopt}
\end{equation}
with $\eta_N$ defined in \eq{eq:defofdimless}, \eq{eq:etaNg}. The
respective flow diagram is given in Fig.~\ref{fig:fullback}.
The flow equations \eq{eq:flowbackZNopt},\eq{eq:flowbackLaopt} admit
an attractive UV fixed point at
\begin{equation}\label{eq:FPfullback}
({g_{N}}_*,\lambda_{*}) =(0.966,0.132)\,, 
\end{equation}
and a repulsive Gaussian fixed point at the origin,
$({g_{N}}_*,\lambda_{*}) =(0,0)$, see also Fig.~\ref{fig:fullback}.   

\begin{figure}[t]
   \includegraphics[width=.88\columnwidth]{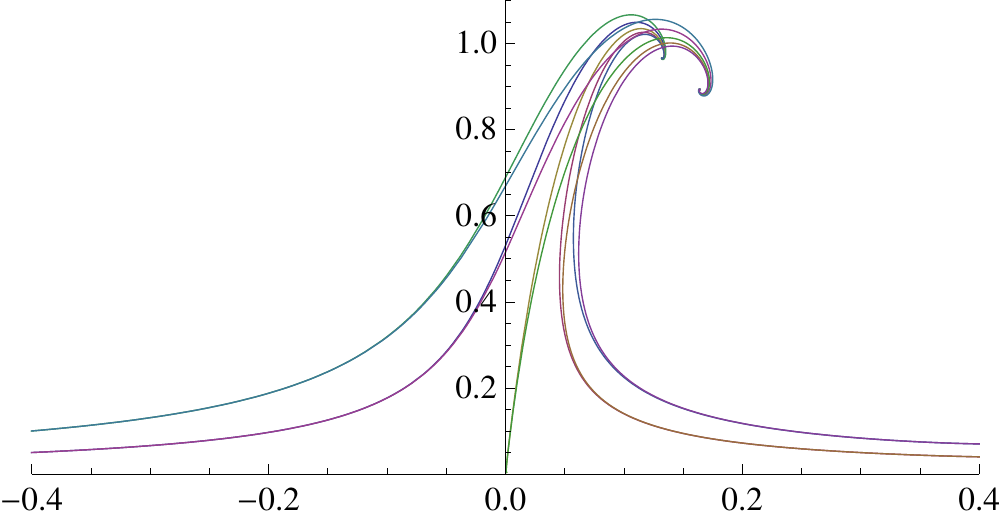}
\put(-130,105){\large $g_k$}
\put(2,5){\large $\lambda_k$}    
\caption{Phase diagrams in the background field approximation from the
  flows \eq{eq:FPback}, \eq{eq:FPfullback} in
  Section~\ref{sec:backgroundflow} with FP $(0.893,0.164)$, and from
  the flows \eq{eq:flowbackZNopt} and \eq{eq:flowbackLaopt} with FP
  $(0.966,0.132)$.}
\label{fig:backfullback}\end{figure}
Interestingly, the flows \eq{eq:flowbackZN}, \eq{eq:flowbackLa} do not
agree with those in Section~\ref{sec:backgroundflow}, equations
\eq{eq:backg}, \eq{eq:backlambda}. This leads to a different phase
diagram, see Fig.~\ref{fig:backfullback}, and different fixed point
values, \eq{eq:FPback} and \eq{eq:FPfullback}. We emphasise that the
position of the fixed points are not physical observables and depend
on the parameterisation of the theory. Indeed the differences depicted
in Fig.~\ref{fig:backfullback} are small and are comparable to
differences obtained by varying the regulators. The latter variation
tests the stability of the approximation at hand as well as the
reparameterisation independence.  Moreover, the differences are fully
explained by an additional approximation made in the standard
background field approximation which is not present in the flows
\eq{eq:flowbackZN}, \eq{eq:flowbackLa}. In turn, the flows in
Section~\ref{sec:backgroundflow} were constructed within the same
approximation commonly used in the background field approach, see
e.g.\ the reviews \cite{Reuter:2007rv,Percacci:2007sz,Litim:2011cp,Reuter:2012id}
and literature therein. The only difference to the standard
approximation in the background field approach in
Section~\ref{sec:backgroundflow} is the use of the covariant
derivatives in the two point functions.  As discussed in
Section~\ref{sec:backgroundflow} the two flows agreed in four
dimensions. The difference to the present flows occurs in
$I_{\lambda,N}^{(2)}$, the coefficients of $\eta_N$ in the flow. It
has but nothing to do with the difference between geometrical flow and
background flow but relates to an additional approximation usually
applied in the latter. For computational simplicity the wave function
renormalisation $Z_N$ has also been applied to all terms in the
effective action, also to the gauge fixing term with
$Z_\alpha=Z_N$. The latter term, however, does not run with
$Z_N$. Indeed, any flow of the gauge fixing term only signals the
breaking of diffeomorphism invariance. In the standard background
field approach such a flow in induced by the cut-off term but does not
agree with the flow of $Z_N$. In turn, in the present
diffeomorphism-invariant setting $Z_\alpha$ does not flow. This
singles out the flows \eq{eq:flowbackZN}, \eq{eq:flowbackLa} as the
correct implementation of the background field approximation in the
present setting. Due to the formal equivalence of both approaches in
Landau-DeWitt gauge it is suggestive that one also should set
$Z_\alpha=1$ in the standard background field flow. In conclusion the
geometrical flow in the background field approximation agrees in four
dimensions with the standard background field flow in the standard
background field approximation within the Einstein-Hilbert truncation together
with the above treatment of the gauge fixing term.

\section{Dynamical Phase diagram \& Fixed points} \label{sec:geoflucresults} 

Now we proceed to the full system including also the flow of
fluctuation couplings $(g_N,\lambda)$. First we remark that for
general regulators the flows \eq{eq:flowbackZN}, \eq{eq:flowbackLa} do
neither have the form \eq{eq:flowbar}, nor does the flow for
$(g_N,\lambda)$ have the form \eq{eq:flowfluc}. This is only achieved
for regulators leading to threshold functions $\Phi$ and $\tilde\Phi$, 
see Appendix~\ref{app:Is}, \eq{eq:thres}, that satisfy  
\begin{equation}\label{eq:Iconstraint} 
   \Phi^1_{\0{d-2}{2}} /\tilde\Phi^1_{\0{d-2}{2}} = \frac{d}{2}\,,\qquad  
   {\Phi^2_{\0{d}{2}}}/{\tilde\Phi^2_{\0{d}{2}}}= \frac{d+2}{2}\,, 
\end{equation}
\Eq{eq:Iconstraint} holds for the optimised regulator,
see \eq{eq:optrel}. Note that the latter was used to derive
\eq{eq:flowbar}, \eq{eq:flowfluc} in the first place. Evidently,
\eq{eq:Iconstraint} holds for a larger class of regulators as it
comprises only two integral constraints on a given regulator. However,
if one improves the current approximation, further constraints arise,
leading to \eq{eq:optrelgen} for $n=d/2$ and $(d-2)/2$. This uniquely
singles-out the optimised regulator. For general regulators one might
compute all the necessary coefficient functions. However, it is more
convenient to use the approximation \eq{eq:Iconstraint} on the basis
of explicitly computing $F^{(2)}$. Within this approximation it is
easily possible to map the known background results in the literature
to the full flow where one distinguishes between dynamical fluctuation
fields and background fields.

\begin{figure}[t]
  \includegraphics[width=.85\columnwidth]{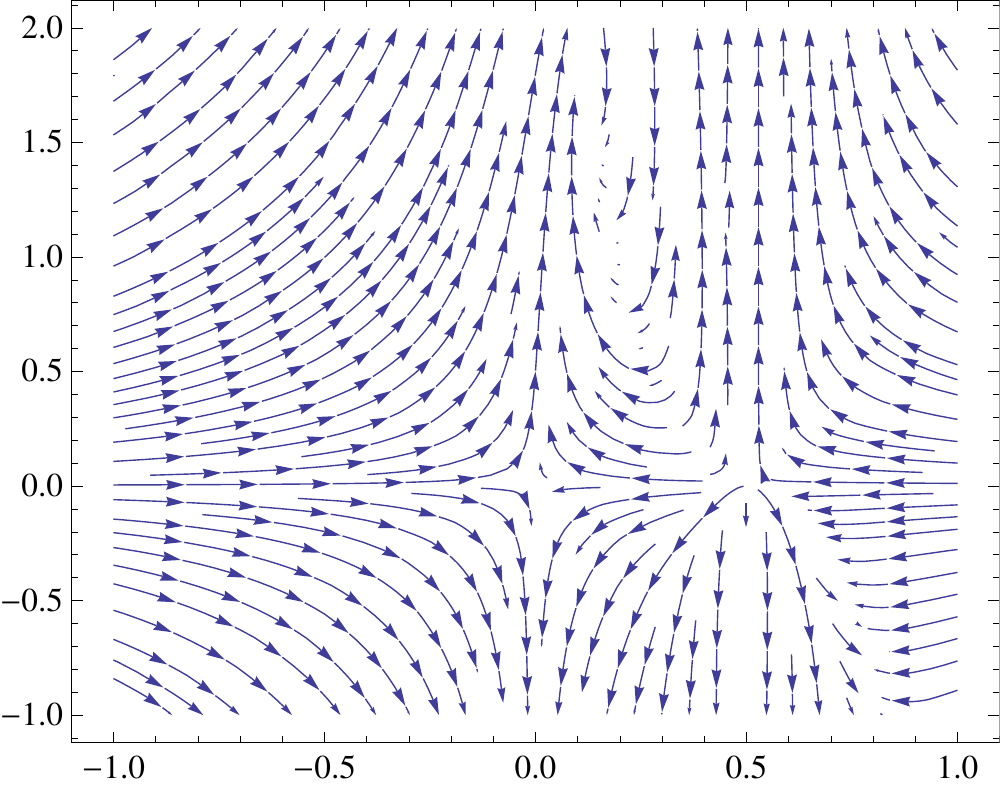}
\put(-220,150){\large $g_k$}
\put(5,5){\large $\lambda_k$} 
  \caption{Phase portrait of the dynamical flow in terms of the vector
  field $(\partial_t g_N,\partial_t \lambda)$}
\label{fig:vectorfull}\end{figure}
We continue with our analysis of the optimised flow. The flow equations of the
dynamical fluctuation couplings $(g_N,\lambda)$ follow from
\eq{eq:flowfluc}  as
\begin{equation}
\left( \partial_t + (2-d) \right) g_N
 =2 \CI_{N}+2 \CI_{N,1}-
\left(\eta_N+\dot \lambda\partial_\lambda\right) I_{N,-2}
  \,, \label{eq:flowZNfluc} 
\end{equation} 
with $\CI_N$ defined in \eq{eq:Is} and 
\begin{equation}
\left( \partial_t +(2-\eta_N)\right)\lambda =
  2 \CI_\lambda-\left(\eta_N+\dot \lambda\partial_\lambda\right)  I_{\lambda,-2}\,,
 \label{eq:flowLafluc}\end{equation} 
and \eq{eq:Iconstraint} holds. The coefficient functions
$\,I_{\lambda,N}\,$ (with $d=4$) have been already used for the flows
in the background couplings,
\eq{eq:flowbackZNopt}, \eq{eq:flowbackLaopt}. Together with the
right hand sides of \eq{eqflowg} and \eq{eqflowgbar} they 
can be found in Appendix~\ref{app:thresopt}. With the results of
Appendix~\ref{app:thresopt} we arrive at the flow equations for the
dynamical couplings $g_N,\lambda$,
\begin{eqnarray}\nonumber 
\partial_t g_N &=&2 g_N-\frac{g_N^2}{\pi} \frac{\frac{5}{9}+\frac{1}{3}\,(1-2\lambda)+
\frac{5}{9}\frac{25}{24}(1-2\lambda)^2}{ (1-2\lambda)^2 -\frac{g_N}{2\pi}
\left(\frac{5}{9}+\frac{1}{3}{(1-2\lambda)}\right) } \\[1ex]
& & \hspace{-.8cm}+\0{2\partial_t \lambda}{1-2 \lambda} \frac{g_N^2}{2\pi} \frac{\frac{10}{9}+\frac{1}{3}\,(1-
2\lambda)}{ (1-2\lambda)^2 -\frac{g_N}{2\pi}
\left(\frac{5}{9}+\frac{1}{3}{(1-2\lambda)}\right) }\,, \label{eq:dynflowg} 
\end{eqnarray}
and 
\begin{equation}
  \partial_t \lambda =\0{-(2-\eta_N) \lambda+(2-\eta_N) 
    \frac{g_N}{4\pi}\frac{1}{1-2\lambda} 
  -\frac{g_N}{3\pi}}{1+
\0{g_N}{2\pi}\0{1}{(1-2\lambda)^2}}\,, \label{eq:dynflowlambda} 
\end{equation}
with $\eta_N$ defined in \eq{eq:defofdimless}, \eq{eq:etaNg}.  The
flow equations \eq{eq:dynflowg} and \eq{eq:dynflowlambda} describe the
phase diagram of quantum gravity in the extended Einstein-Hilbert
truncation in terms of the dynamical couplings $g_N,\lambda$. The
vector fields of the corresponding $\beta$-functions are depicted in
Fig.~\ref{fig:vectorfull}. In
comparison to the standard background flows
\eq{eq:backg}, \eq{eq:backlambda} and the geometrical background flows
\eq{eq:flowbackZNopt}, \eq{eq:flowbackLaopt} they contain a further
resummation. The related terms are given with the second line in
\eq{eq:dynflowg} and the non-trivial denominator in
\eq{eq:dynflowlambda}. They are related to scale derivatives
$\partial_t \lambda$ and can be understood in terms of standard 2PI
and hard thermal loop resummations of the self energy or mass in
quantum field theory, see \cite{Blaizot:2010zx} for the FRG
implementation.  This new, additional resummation removes the infrared
singularity in the flows at $\lambda=1/2$ present in the background
field flows but introduces new repulsive singularities in the flow of
the Newton constant $g_N$. This is very reminiscent of the screening of the infrared
singularity in thermal theories and is discussed in
Section~\ref{subsec:IR}. 

\begin{figure}[t]
   \includegraphics[width=.88\columnwidth]{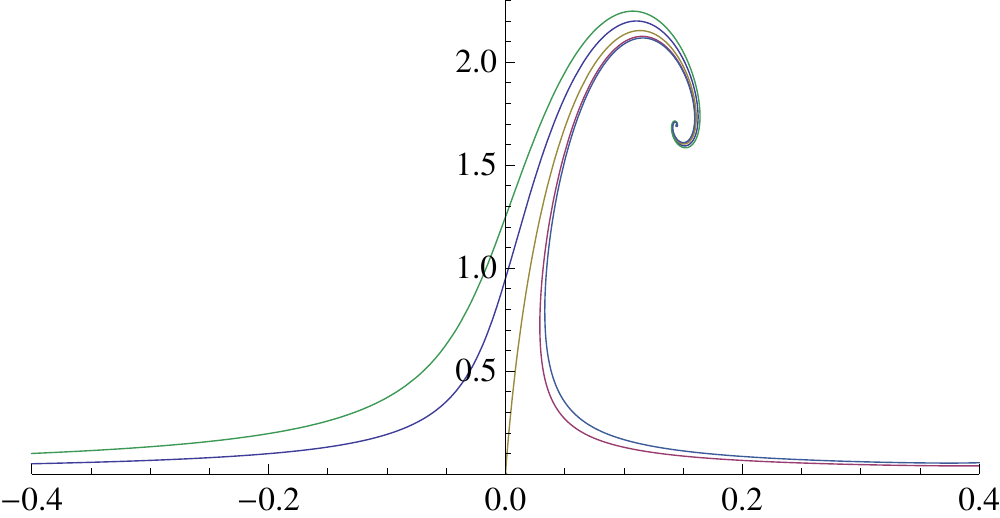}
\put(-130,105){\large $g_k$}
\put(2,5){\large $\lambda_k$}  
\caption{Phase diagram for the dynamical couplings, \eq{eq:dynflowg}
  and \eq{eq:dynflowlambda}, with an UV fixed point
  $({g_N}_*,\lambda_{*})=(1.692, 0.144)$, and the repulsive
  perturbative fixed point ${\rm FP}_{\rm rep}=(0,0)$.}
\label{fig:dynflow}\end{figure}
The phase portrait in Fig.~\ref{fig:vectorfull} shows an attractive UV
fixed point as well as a repulsive Gaussian fixed point at the origin
in analogy to the background field approximations.  It also shows an
attractive IR fixed point at $g_N=0$ and $\lambda=1/2$ as well as
repulsive lines emanating from the IR fixed point.

\subsection{UV fixed point}\label{subsec:UV}
\begin{figure}[t]
   \includegraphics[width=.88\columnwidth]{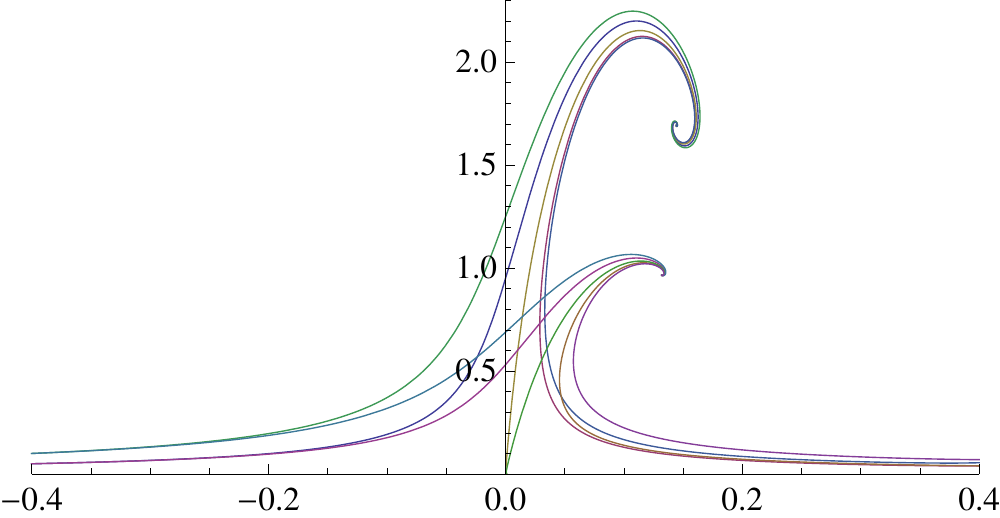}
\put(-130,105){\large $g_k$}
\put(2,5){\large $\lambda_k$}  
   \caption{Phase diagrams in the background field approximation from
     the flows \eq{eq:flowbackZNopt} and \eq{eq:flowbackLaopt} with
     the UV fixed point $(0.966,0.132)$, and from the dynamical flows
     \eq{eq:dynflowg} and \eq{eq:dynflowlambda} with the UV fixed point 
     $(1.692, 0.144)$.}
\label{fig:dynfullback}\end{figure}

The second line in \eq{eq:dynflowg} drops out at a fixed point. The
flow equations \eq{eq:dynflowg},\eq{eq:dynflowlambda} admit an
attractive UV fixed point at 
\begin{equation}\label{eq:dynFP}
\rm FP^{\ }_{\rm \tiny UV}=({g_{N}}_*,\lambda_{*}) =(1.692, 0.144)\,, 
\end{equation}
and a repulsive Gaussian fixed point at the origin,
${\rm FP}_{\rm rep}=({g_{N}}_*,\,\lambda_{*}) =(0,0)$, see also Fig.~\ref{fig:dynflow}. 
The flow diagram and the fixed point differs from that in the
background field approximation depicted in
Fig.~\ref{fig:fullback}. A comparison of the respective phase diagrams is depicted
in Fig.~\ref{fig:dynfullback}. 
\begin{table}[b]
\caption{Fixed Points}\label{tab:FPs}
\centering
\begin{tabular}{|c|c|c|c|}
\hline
\quad  Type of flow \,\,\quad & ${g_N}_*$ & $\lambda_*$ & $\,\,{g_N}_*\,\times\,\lambda_*\,\,$ \\
\hline
\quad Background  \quad  &  \quad  0.893 \quad      &  \quad     0.164  \quad     &   \quad    0.146  \\
\hline
\quad Improved background  \quad  &  \quad  0.966 \quad      &  \quad     0.132  \quad     &   \quad    0.128  \\
\hline 
\quad Dynamical  \quad  &  \quad  1.692 \quad      &  \quad     0.144  \quad     &   \quad    0.244  \\
\hline
\quad Bimetric  \quad  &  \quad  1.055 \quad      &  \quad     0.222  \quad     &   \quad    0.234  \\
\hline
\end{tabular}
\end{table} 
The difference of the two phase diagrams in Fig.~\ref{fig:dynfullback}
is qualitatively different from that between the two background field
approximations discussed before. It is not comparable to differences
obtained by varying the regulator. This is best seen by studying the
product of Newton constant and cosmological constant which is
significantly reduced for the dynamical flow in comparison to the
background field approximation, see Table~\ref{tab:FPs}. We have also
included a comparison to fixed points derived within the bimetric
flows studied in \cite{Manrique:2010am}. While the positions of the
fixed points for the dynamical and bimetric flows are quite different,
the invariant product ${g_N}_*\lambda_*$ only deviated by a few precent. 

\begin{table}[h]
\caption{Stability matrices}
\centering
\begin{tabular}{|c|c|c|c|}
\hline
\quad Flows \,\,\quad & Stability matrix & Eigenvalues \\
\hline
 \, Background  \quad  &    $\left(\begin{array} {cc}
                                             -2.46 & -10.52  \\
                                              0.71 & -1.61  
                                        \end{array}\right)$    &   $\begin{array}{c}
                                                                    -2.03 + 2.69i  \\
                                                                    -2.03 - 2.69i   
                                                                    \end{array} $  \\
\hline
 \,  Improved backgr.  \quad  &  $\left(\begin{array} {cc}
                                             -2.59 & -9.99  \\
                                              0.47 & -2.01  
                                            \end{array}\right)$ &   $\begin{array}{c}
                                                                    -2.30 - 2.16i  \\
                                                                    -2.30 - 2.16i    
                                                                    \end{array} $  \\   
\hline              
 \,  Dynamical \quad  &  \quad  $\left(\begin{array} {cc}
                                             -1.94 & -27.9  \\
                                         0.26 & -0.74 
                                            \end{array}\right)$ \quad      &  $\begin{array}{c}
                                                                              -1.34 + 2.61 i    \\
                                                                               -1.34 - 2.61 i   
                                                                               \end{array} $      \\
\hline
\end{tabular}
\label{tab:stab}\end{table} 
The stability matrices are displayed in Table~\ref{tab:stab}, the
bending around the fixed point, which is introduced by the imaginary
part of the eigenvalues, is reduced from the standard background field
approximation to the improved one. For the dynamical flows, the
bending is stronger which comes from the $\partial_t\lambda$-terms 
in the flows. Without these terms the bending is even reduced further 
in comparison to the improved background flows.  
This leads to a far smaller bending of the phase diagram about the
fixed points, see Fig.~\ref{fig:dynfullback}.

\subsection{IR fixed points \& phase diagram}\label{subsec:IR}

\begin{figure}[t]
   \includegraphics[width=.88\columnwidth]{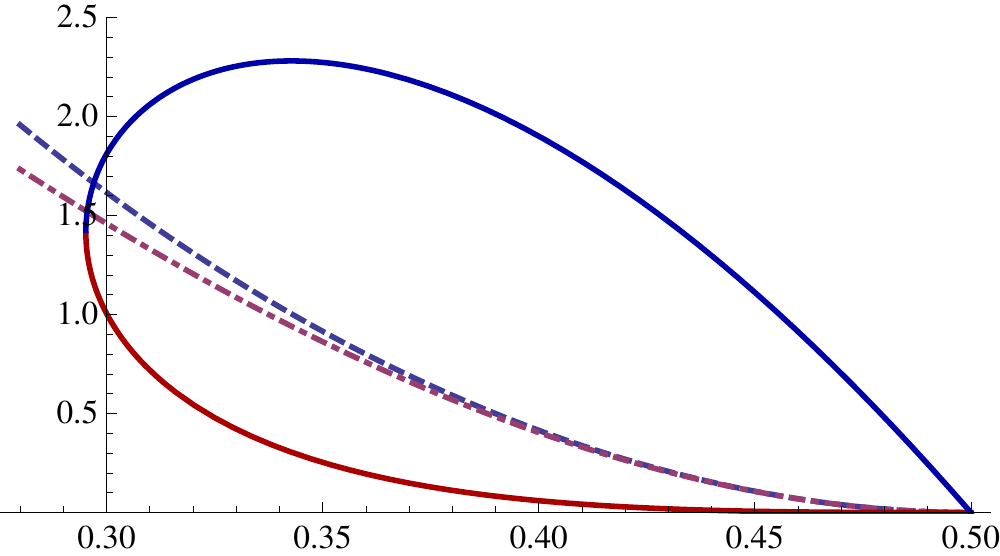}
\put(-225,105){\large $g_k$}
\put(2,5){\large $\lambda_k$}    
\caption{Singular lines for the infrared-directed flow $(-\partial_t
  g_N,-\partial_t \lambda)$. The dashed, dotdashed and full lines are
  the singular lines for the standard background, the improved
  background and the dynamical flow respectively. }
\label{fig:singular}\end{figure}

Finally, we would like to discuss the infrared behaviour of the full
dynamical flows \eq{eq:dynflowg},\eq{eq:dynflowlambda}. A similar
investigation was first done in \cite{ContrerasLitim} within the
background field approach. The resummation due to the
$\partial_t\lambda$-terms screens the singularity in the propagators
at $1-2\lambda=0$ similarly to the screening of thermal infrared
singularities via thermal resummations. However, the resummations
related to $\eta_N$ on the right hand side of the flow lead to
singular lines in the flow diagram, where both flows, $\partial_t g_N$
and $\partial_t \lambda$, exhibit poles at
\begin{eqnarray}\nonumber
  \hspace{-.3cm}g^\pm_{N}&=&\015\pi \Bigl[ (-(1-2\lambda) 
  +22 \lambda(1-2 \lambda) \\[1ex]
  &&\pm\sqrt{ \lambda(1-2 \lambda) ^2 (179-676 
    \lambda +236\lambda^2)}\Bigr] 
  \,. 
 \label{eq:gsing}\end{eqnarray}
The singular lines terminate at $(g_N,\lambda)=(1.414, 0.295)$, 
and $(g_N,\lambda)=(0,1/2)$. Such singular lines are as well present
in the background field flows as they also include resummations
related to $\eta_N$. Here we concentrate on positive cosmological
constant where the background flows exhibit singular lines that
terminates at $\lambda=4/3$ (improved background flow) or at
$\lambda=1/30 (9 \pm 4\sqrt{21})$ (standard background flow) and
extend to $\lambda=0$. The singular lines are displayed in
Fig.~\ref{fig:singular}. 

%
%

\begin{figure}[t]
   \includegraphics[width=.88\columnwidth]{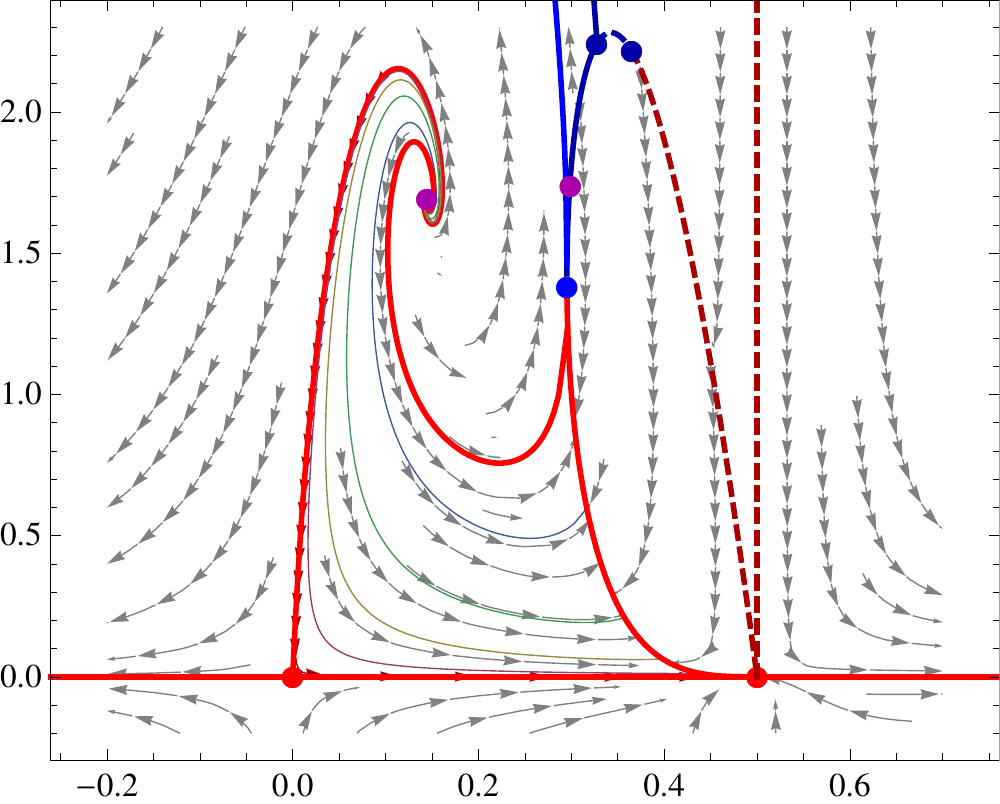}
\put(-180,80){\Large $\rm Ib$}
\put(-130,55){\Large $\rm Ia$}
\put(-225,160){\large $g_k$}
\put(2,5){\large $\lambda_k$}   
\caption{Full phase diagram for the dynamical couplings including the
  repulsive perturbative fixed point ${\rm FP}_{\rm rep}=(0,0)$ and ${\rm FP}_{\rm
    IR,1}=(1/2,0)$. The red boundary lines show the separatrices.}
\label{fig:dynstable}\end{figure}
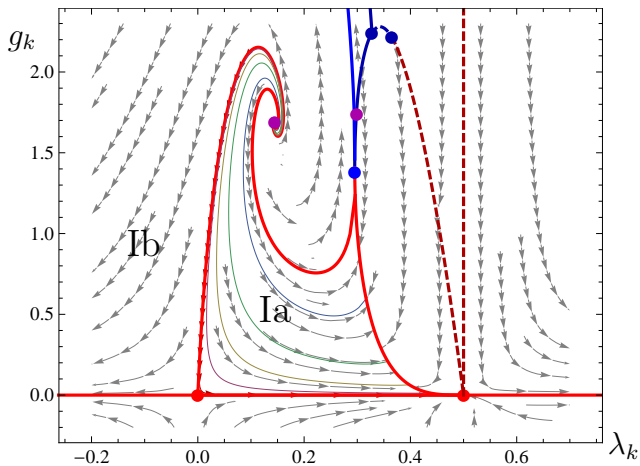

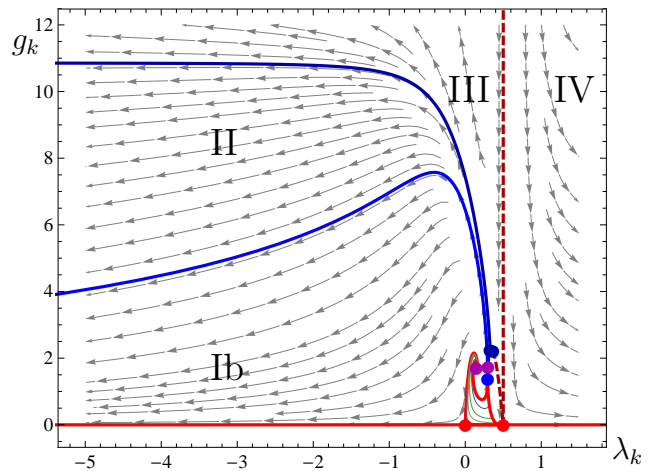
\begin{figure}[t]
  \includegraphics[width=.88\columnwidth]{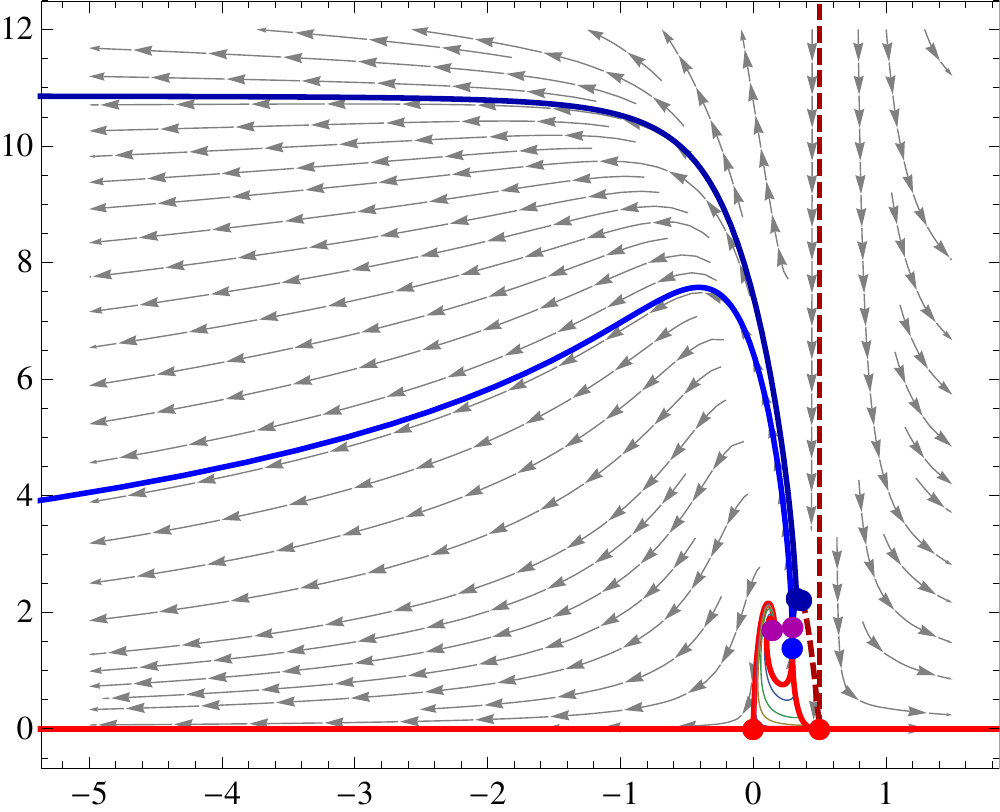}
\put(-20,140){\Large $\rm IV$}
\put(-60,140){\Large $\rm III$}
\put(-150,120){\Large $\rm II$}
\put(-150,35){\Large $\rm Ib$}
\put(-225,160){\large $g_k$}
\put(2,5){\large $\lambda_k$}   
\caption{Full phase diagram for the dynamical couplings including  the
  repulsive perturbative fixed point ${\rm FP}_{\rm IR,0}=(0,0)$, and the 
attractive IR fixed points ${\rm FP}_{\rm
    IR,1}=(1/2,0)$ and ${\rm FP}_{\rm
    IR,2}=(-\infty,0)$.}
\label{fig:dynstableglobal}\end{figure}

All flows go through the point $(\lambda,g_N)=(1/2,0)$, for the
background flows, however, the $\lambda$ axis is tangential to the
singular lines at $(1/2,0)$. For the dynamical flow the area under the
singular line is restricted by $g_-$ in \eq{eq:gsing}. All flows in
this area hit the singular line at some point: let us assume that
there are flows that go to $(1/2,0)$ without hitting the singular
line. Then, in the vicinity of $(1/2,0)$ we expand $g^-_{N}$ about
$\lambda=1/2$, leading to 
\begin{eqnarray}
g^-_{N}&=& \095\pi (1-2 \lambda)^3+O[(1-2 \lambda)^4]\,.
 \label{eq:gsingexpand}\end{eqnarray}
This entails that the dynamical flows, \eq{eq:dynflowg} and
\eq{eq:dynflowlambda}, reduce to
\begin{equation}\label{eq:dynreduced}
\partial_t g_N= 2 g_N(1+ \partial_t \lambda)\,,\qquad 
\partial_t \lambda=-(2-\eta_N)\lambda\,, 
\end{equation}
which implies that $\partial_t g_N =-4 g_N/(1-2 \lambda)<0$ in leading order. Hence,
with \eq{eq:dynreduced} we conclude that all trajectories in the
vicinity of $\lambda=1/2$ hit the singular lines. On the lower
singular line given by $g_N^-$ the infrared-directed flows diverge but
point towards the singular line. Hence this line is infrared
stable. Moreover, there is a finite net flow on this singular line
which comes from the sum of the (singular) flows taken at both sides
of the singular line: we define unit infrared-directed tangential
vectors $\hat e(\lambda,g_N)$ to given trajectories with
\begin{equation}\label{eq:ebeta}
\hat e(\lambda,g_N)= \0{\vec\beta}{\|\vec \beta\|}\,,\qquad 
\vec \beta(\lambda,g_N)=-( \partial_t\lambda\,,\, \partial_t g_N) \,,
\end{equation} 
and the tangential vector on the singular line as a function of
$\lambda$ is given by
\begin{equation}\label{eq:eminus}
\hat e_\pm(\lambda)=\0{1}{\sqrt{1+(\partial_\lambda g^\pm_{N})^2}}
( 1\,,\, \partial_\lambda g^\pm_{N}) \,.  
\end{equation}
The $e_\pm$ point towards $(1/2,0)$ along the respective singular line
given by $g_N^\pm$. The corresponding orthogonal boundary vectors
$\hat e^\bot_{\pm}$, directed away from the region bounded by the
singular line, are given by
\begin{equation}\label{eq:ebot}
\hat e^\bot_{\pm}(\lambda)=\pm \0{1}{\sqrt{1+(\partial_\lambda g^\pm_{N})^2}}
( -\partial_\lambda g^\pm_{N}\,,\,1) \,.  
\end{equation}
The above definitions allow us to define the finite net flow vector
$\vec \beta_{\rm net}$ on the singular line and the corresponding unit
flow vector $\hat \beta_{\rm net}$ are given by
\begin{eqnarray}\nonumber 
  \vec \beta_{\rm net}^\pm &=& 
\lim_{\epsilon\to 0}\012 \Bigl[\vec\beta\bigl( (\lambda,g^\pm_{N})+\epsilon 
\hat e_\pm(\lambda)\bigr)\\[1ex]\nonumber 
&&\hspace{1.5cm}+\vec\beta\bigl( (\lambda,g^\pm_{N})-\epsilon 
\hat e_\pm(\lambda)\bigr)\Bigr]\,, \\[1ex]
 \hat \beta_{\rm net} &=&\0{\vec 
\beta_{\rm net}}{\|\vec \beta_{\rm net}\|}\,.
\label{eq:betanet}\end{eqnarray} 
%
\begin{figure}[t]
   \includegraphics[width=.80\columnwidth]{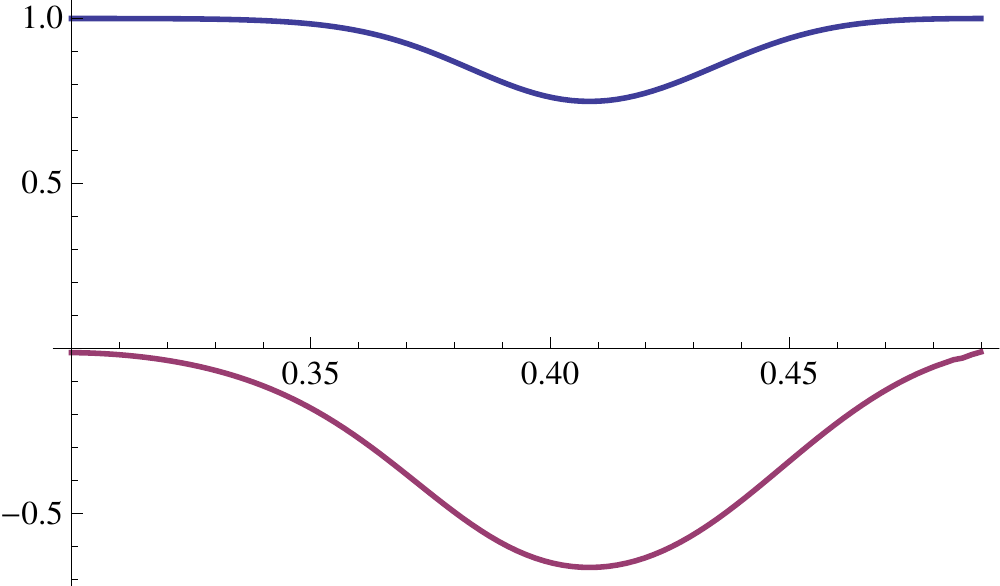}
\put(-178,100){$\hat\beta^-_{\rm net}\hat e_-$}
\put(-178,32){$\hat\beta^-_{\rm net}\hat e_-^\bot $}
\put(-15,30){\large $\lambda$}
\caption{Coefficients of the unit net flow on the singular line given by $g_N^-$,  
\eq{eq:gsing}, directed towards $(1/2,0)$. }
\label{fig:betanet}\end{figure}
%
%
\begin{figure}[t]
   \includegraphics[width=.80\columnwidth]{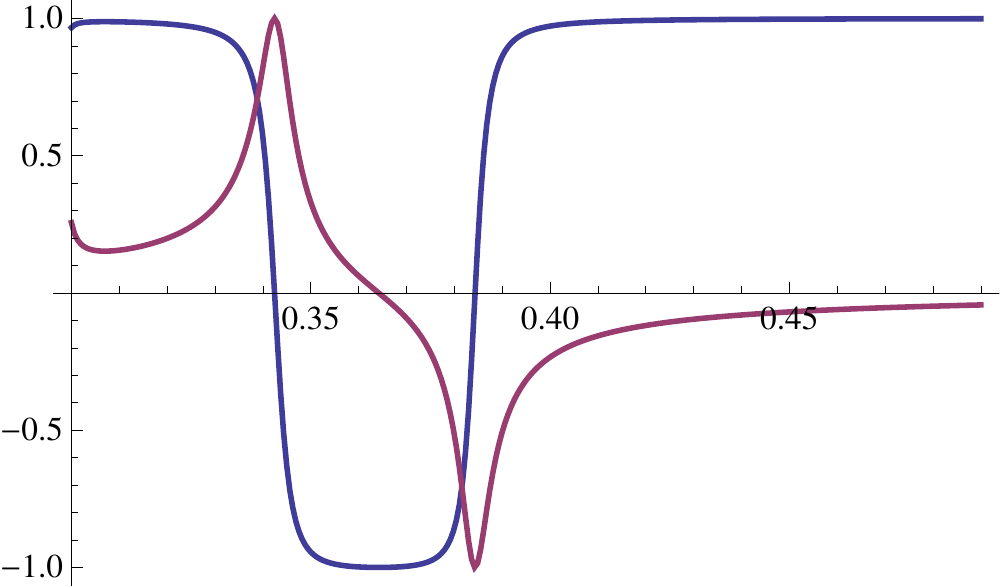}
\put(-178,117){$\hat \beta^+_{\rm net}\hat e_+$}
\put(-178,82){$\hat \beta^+_{\rm net}\hat e_+^\bot $}
\put(-15,40){\large $\lambda$}
\caption{Coefficients of the unit net flow on the singular line given by $g_N^+$,  
\eq{eq:gsing}, directed towards $(1/2,0)$. }
\label{fig:betanetplus}\end{figure}
%
%
\begin{figure}[t]
   \includegraphics[width=.80\columnwidth]{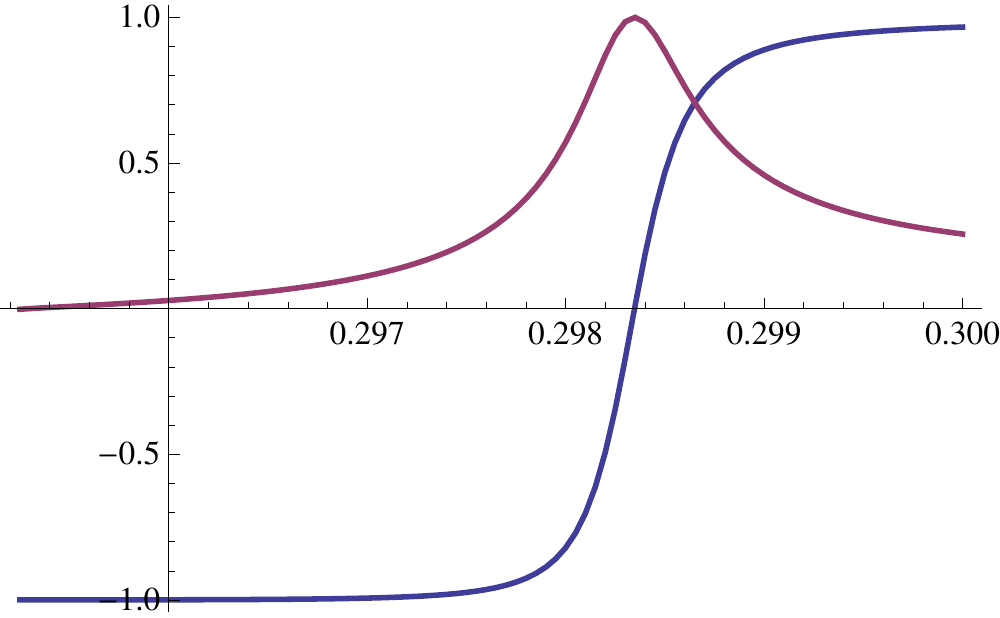}
\put(-195,10){$\hat \beta^+_{\rm net}\hat e_+$}
\put(-195,72){$\hat \beta^+_{\rm net}\hat e_+^\bot $}
\put(-15,40){\large $\lambda$}
\caption{Coefficients of the unit net flow on the singular line given by $g_N^+$,  
\eq{eq:gsing}, directed towards $(1/2,0)$ in the vicinity of the turning point. }
\label{fig:betanetplus1}\end{figure}
%
In case of the infrared stable part of the singular line there is a
flow along the singular line with the strength $|\vec \beta_{\rm
  net}\cdot \hat e |$. The direction is given by the sign of $\cos
\theta= \hat \beta_{\rm net} \cdot \hat e$ where $\theta$ is the angle
between $\vec \beta_{\rm net}$ and $\hat e$. This is plotted in
Figs.~\ref{fig:betanet}, \ref{fig:betanetplus} and
\ref{fig:betanetplus1}. In these figures we also plot $\cos
\theta^\bot = \hat \beta\cdot \hat e^\bot$ which encodes the
information, whether the net flow is directed into the region bounded
by the singular line or away from it. The full phase portrait is depicted in
Fig.~\ref{fig:dynstable} and Fig.~\ref{fig:dynstableglobal}. 

Most importantly, the projected net flow $\hat \beta^-_{\rm net}\cdot
\hat e_-$ on $g^-_N$ is directed towards $(1/2,0)$. Hence the lower
singular line is fully infrared stable, see Fig.\ref{fig:betanet}. The
infrared attractive point $(1/2,0)$ is reached after a finite flow
time at the cut-off scale $k_0$. For $k<k_0$ the flows are trivial,
\begin{equation}\label{eq:trivialflow}
\partial_t g_N=0\,,\quad \partial_t\lambda = -2\lambda\quad \rightarrow\quad 
g_N=0\,,\quad \lambda = \0{k_0^2}{2 k^2}\,.
\end{equation}
\Eq{eq:trivialflow} reflects a trivial fixed point of a free massive
theory. Note that this interpretation should be taken with caution due
to the singularities. This concerns in particular the quantitative
results, such as $g_N\lambda=0$ in the infrared. In summary we are led
to the UV-IR stable region~Ia with the UV-attractive fixed point $\rm
FP^{\ }_{\rm \tiny UV}=(1.692, 0.144)$ and the IR-attractive fixed
point ${\rm FP}_{\rm \tiny IR,1}=(0, 1/2)$, see
Fig.~\ref{fig:dynstable}. In turn, at the turning point of the
singular line, $g_+=g_-$ with vertical tangential vector,
\begin{equation}\label{eq:turning} 
\lambda=\0{(169-60 \sqrt{5})}{118} \,,\quad  g_N^+=g_N^-
=\0{120\pi (301 \sqrt{5}-660)}{ 3481}\,,
\end{equation}
the sign of $\cos\theta^\bot$ turns positive and the net flow is
directed away from the singular line. As this is also true for the
full $\beta$ functions, the singular line gets infrared instable (and
ultraviolet stable), see Fig.~\ref{fig:betanetplus1}. In any case, no
flow from the UV fixed point can reach this part of the singular
line. This defines a separatrix from the UV fixed point to the turning
point $(0.295, 1.414)$ of the singular line with $g_+=g_-$, see
\eq{eq:turning}. Flows below this separatrix are driven towards the
attractive infrared fixed point $(1/2,0)$, flows above the separatrix
are driven towards the attractive infrared fixed point
$(-\infty,0)$. In summary this leads the UV-IR stable region~I with
the UV attractive fixed point ${\rm FP}_{\rm \tiny UV}=(1.692,
0.144)$, the repulsive fixed point ${\rm FP}_{\rm \tiny rep}=(0, 0)$
and the two IR attractive fixed points ${\rm FP}_{\rm \tiny
  IR,1}=(1/2,0)$ and ${\rm FP}_{\rm \tiny IR,2}= (-\infty,0)$. Similar
results have been obtained in the standard background field 
approach in \cite{ContrerasLitim}. 

%
The region II in Fig.~\ref{fig:dynstableglobal} can be accessed from the infrared fixed point ${\rm
  FP}_{\rm \tiny IR,2}$. UV flows in this region hit the singular line
between $0.295<\lambda<0.327$, depicted by the upper light blue and
left dark blue dots in Fig.~\ref{fig:dynstable}. We remark that this
part of $g_N^+$ is ultraviolet attractive and UV flows are driven
towards $\lambda=0.298$ on the singular line. Accordingly there is a
potential further UV attractive fixed point at ${\rm FP}_{\rm \tiny
  UV,2}=(0.298,1.738)$, depicted with a violet dot in
Fig.~\ref{fig:dynstable}. This singles out the second IR-UV attractive
region~II. Note that this may very likely be an artefact of the
approximation. Still it is worth further consideration.

The regions III and IV cannot be accessed from the UV fixed points nor
do flows in the regions III and IV reach the infrared fixed points
${\rm FP}_{\rm \tiny IR,1}$ or ${\rm FP}_{\rm \tiny IR,2}$. Infrared
flows in this region are driven towards $(\infty,0)$ or
$(\infty,\infty)$. 

The same analysis can be made for the standard background and improved
background approximation. We only mention that there exists regions
similar to region Ia, and flows are directed towards the endpoint
$(1/2,0)$ for $\lambda>0.4635$ (improved background) and for
$\lambda>0.4637$ (standard background). Interestingly for both flows
the scalar product $\vec \beta_{\rm net}\cdot \hat e_{-}$ is also positive
for $\lambda<0.341$ (improved background) and for
$\lambda>0.335$ (standard background). This leads to a further
infrared stable point at $(\lambda,g_N)=(0.341, 0.966)$ and
$(\lambda,g_N)=(0.335, 1.108)$ respectively.

In summary all flows exhibit a region which is ultraviolet and
infrared stable. This is depicted for the dynamical flows in
Figs.~\ref{fig:dynstable},\ref{fig:dynstableglobal}. At the IR fixed
point $FP_{\rm IR,2}=(1/2,0)$ the product $g_N\lambda=0$. Hence,
$\lambda$ vanishes in terms of $g_N$ in the infrared. Note however,
that $\lambda$ is simply a parameter in the propagator of the
fluctuation field $h$ and its interpretation as the cosmological
constant is not straightforward. 
%
%

\subsection{Matrix elements \& observables}\label{subsec:physob}
It is left to determine physics observables such as the strength of
the gravitational interaction measure in experiments. Here we only
present the flow equations for the physical Newton coupling and
cosmological constant and discuss the consistency of the results with
the analysis made so far. Note first, that the dynamical couplings
$g_N,\lambda$ are only indirectly related to physics observables. This
is a property the geometrical approach shares with the standard
background field approach to quantum field theory. However, we have
also seen, that the background couplings are sensitive to the
regulator. This holds in particular in the scaling regions: the
regulators have been chosen such that they show the same (singular)
scaling as the corresponding two-point functions $\Gamma^{(2)}$. Such
regulators are called RG-adapted,
\cite{Pawlowski:2005xe,Pawlowski:2001df}, or spectrally adjusted,
\cite{Gies:2002af}. The effective action satisfies the RG and scaling
equations of the underlying full theory at vanishing cut-off, see
\cite{Pawlowski:2005xe,Pawlowski:2001df}. This property facilitates
the access to scaling regions and relates to a partial optimisation of
the flow, \cite{Pawlowski:2005xe}, but complicates the extraction of
the physical part of the background field correlation functions.  It
has been shown in the standard background field approach that the
regulator-induced terms can even change the sign of the
$\beta$-functions, see \cite{Litim:2002ce}, in the context of gravity
this has been discussed in \cite{Folkerts:2011jz}.

In the geometrical approach it is the Nielsen identity \eq{eq:Nielsen} that
controls the difference between background field dependence and
fluctuation field dependence, the right hand side being the term
stemming from the regulator, see
\cite{Pawlowski:2003sk,Pawlowski:2005xe}. Hence, at vanishing
fluctuation field $h=0$ the physical background field dependence is
comprised in the standard Nielsen identity
\begin{equation}\label{eq:Nielsenphys}
\left.  \Gamma_{k,i}\right|_{\rm phys}[h=0]=-\Gamma_{k,a}\langle 
\hat h^a{}_{;i}\rangle[h=0]\,. 
\end{equation}
\Eq{eq:Nielsenphys} entails that at $h=0$ we can identity the
$h$-derivatives with the $\bar g$-derivatives up to sub-leading order.
In other words, the physical part of the background couplings, $\bar
g_{N,\rm phys} , \bar\lambda_{\rm phys}$ have $\beta$-functions
similar to that of the dynamical couplings. Note that this argument
fully works the infrared where the regulator tends to zero and the
sub-leading terms are small. It has to be taken with caution for large
regulators. Thus we shall only discuss the infrared region with
$\lambda>1/2$: neglecting the sub-leading terms we arrive at the flow
for the physical part of the background Newton constant,
\begin{eqnarray}
\0{1}{\bar g_{N,\rm phys}}\left( \partial_t +(2-d)\right)\bar g_{N,\rm phys} = 
\0{1}{g_N}F_g(g_N,\lambda)\,,  \label{eq:barZNphys}
\end{eqnarray} 
and the physical part of the 
cosmological constant $\bar\lambda$
\begin{equation}\label{eq:barlambdaphys}
\0{1}{\bar g_{N,\rm phys}}\left( \partial_t +(2-\bar\eta_N)\right)\bar
  \lambda_{\rm phys} =\0{1}{g_N}  F_\lambda(g_N,\lambda)\,.   
\end{equation} 
For the optimised flows the right hand sides $F_g$ and $F_\lambda$ are
given in \eq{eq:finalflowFg} and \eq{eq:finalflowFl} respectively. If
identifying the physical part of the background couplings with the
dynamical ones, $(\bar g_{N,\rm phys} , \bar\lambda_{\rm phys})=(g_{N}
,\lambda)$, we are led to the full flow of the dynamical couplings,
\eq{eq:dynflowg}, \eq{eq:dynflowlambda}. We also remark that with
\eq{eq:barZNphys},\eq{eq:barlambdaphys} we can derive finite net flows
of $g_{N,\rm phys}$, $\lambda_{\rm phys}$ on the singular lines. There, however, the 
sub-leading terms might not be negligible. 

Here we only consider $\lambda>1/2$ with $k>k_0$. Then the
flows \eq{eq:barZNphys},\eq{eq:barlambdaphys} have to be evaluated for
$g_N=0$, to wit 
\begin{eqnarray}\nonumber 
\hspace{-1cm}\partial_t \bar g_{N,\rm phys} &=&2 \bar g_{N,\rm phys}\,,  \\[1ex]
\partial_t \bar
\lambda_{\rm phys}&=&- 2 \bar \lambda_{\rm phys}+
\0{\bar g_{N,\rm phys}}{6 \pi }\0{1 + 8 \lambda(1-\lambda)}{(1 - 2 \lambda)^2}\,.
\label{eq:asymptflow}\end{eqnarray} 
\Eq{eq:asymptflow} implies that $ \bar g_{N,\rm phys}\propto k^2$ for $k\to 0$. We also have 
$\lambda\propto 1/k^2$ due to \eq{eq:trivialflow} and we arrive at 
\begin{equation}\label{eq:asymptflowfinal}
\partial_t \bar g_{N,\rm phys} = 2 \bar g_{N,\rm phys}\,,\qquad 
\lambda_{\rm phys} =- 2 \bar \lambda_{\rm phys}\,, 
\end{equation}
for $k\to 0$. \Eq{eq:asymptflowfinal} simply provides the dimensional
running of Newton constant and cosmological constant. It implies a
finite product $g_{N,\rm phys}\lambda_{\rm phys}$, the value of which
depends on the initial conditions. We conclude that the phase diagram
of quantum gravity in the current approximation shows UV-IR
stability. In the infrared region we are driven towards classical Einstein gravity.

\section{Summary and outlook}\label{sec:summary} 
We close with a brief survey of our results, more detailed
discussions can be found in the respective sections. In the present
work we have established a fully diffeomorphism-invariant flow for
gravity. This flow has also been
shown to be gauge independent in \cite{Pawlowski:2003sk}. In
Section~\ref{sec:backgroundflow} we have shown that the flow agrees in
the linear approximation with the standard background approximation
for the background field flow in Landau-DeWitt gauge. The latter
approximation also implies an artificial scale-dependence (on the wave
function renormalisation $Z_N$) of the longitudinal degrees of
freedom. Note however, that the scaling of the longitudinal (gauge)
degrees of freedom indeed vanishes identically in the geometrical
approach, whereas it only reflects the deformation of diffeomorphism
invariance in the standard background field approach. We are hence
lead to the same flow diagrams and fixed points, and the UV fixed
point is given by in Table~\ref{tab:FPs}. Beyond the linear
approximation the two flows differ in dimensions others than four but
still agree for four dimensions.  We have also introduced an improved
background field approximation where care is taken of the fact that
the longitudinal gauge direction do not flow. The related fixed point
does not differ significantly from the standard background field
result, see Table~\ref{tab:FPs}.
 
Furthermore, we have introduced the difference between the background
metric and the fluctuation metric. This difference has been evaluated
by means of the Nielsen identity derived in \cite{Pawlowski:2003sk},
for the background approach analogue see
\cite{Pawlowski:2001df,Litim:2002ce,Litim:2002xm,Folkerts:2011jz}. While
the fixed point values of the couplings $(g_N,\lambda)$ have no direct
physical meaning, their dimensionless product $g_N \lambda=G_N\Lambda$
differs considerably from that in the background approximation. It
agrees very well with that in the bimetric background field approach,
e.g.\ \cite{Manrique:2010am,Manrique:2010mq}, see Table~\ref{tab:FPs}. 

We have also discussed the infrared behaviour of quantum gravity in
the present approach. Within the present approximation the flows run
into a singularity at $2\lambda=1$ which signals a pole in the
propagator. We emphasise that $\lambda=\Lambda/k^2$ is the
cosmological constant measured in the cut-off scale. The physical
information is stored in $g_N \lambda$, that is, one measures the
cosmological constant in units of the Newton coupling. There are
further singularities in the $\beta$-functions which are related to the
(incomplete) resummations put forward in the present paper. Still one
can define finite net flows on these singular lines, and hence discuss
the resulting phase diagram. A detailed analysis of the phase diagram
reveals a very rich and interesting structure which is discussed in
detail in Section~\ref{subsec:IR}. Note that the respective results
have to be taken with caution. Taking this into account we
find an infrared stable fixed point at ${\rm FP}_{\rm
  IR}=({g_N}_*,\lambda_*)_{\rm IR}=(0,1/2)$ for the dynamical
couplings. Similarly to the standard background field approach these
dynamical parameters have to be mapped to the physical couplings. This
has been done in the last Section~\ref{subsec:physob} where it has
been shown that in the infrared the theory tends towards classical
Einstein gravity.

In summary the present analysis provides the first results within the
fully diffeomorphism-invariant framework introduced in
\cite{Branchina:2003ek,Pawlowski:2003sk}. Additionally it resolves the
difference between fluctuating field and background metric via the
Nielsen identity \cite{Pawlowski:2003sk}. The results of the present
work further solidify the asymptotic safety scenario for quantum
gravity. A more detailed qualitative analysis also reveals a rich
phase structure of quantum gravity including attractive infrared fixed
points. In the infrared the theory tends towards classical Einstein
gravity.  The quantitative understanding of the full phase diagram of
quantum gravity has to be furthered in more elaborated approximations.

{\em Acknowledgements --} We thank N.~Christiansen, S.~Folkerts,
D.~F.~Litim, M.~Reuter, A.~Rodigast and F.~Saueressig for
discussions. JMP thanks D.~F.~Litim for repeated discussions on the
infrared behaviour of gravity and sharing the unpublished results of
\cite{ContrerasLitim}.

\begin{appendix}

\section{York decomposition}\label{app:york}
In the present work we use the York transverse-traceless
decomposition, first introduced in Section~\ref{sec:geometricalflow}
below \eq{eq:covariantbground}. For more details in the context of
FRG-flows see e.g.\ the reviews
\cite{Reuter:2007rv,Percacci:2007sz,Litim:2011cp} and literature
therein, as well as \cite{donkin}. The York decomposition amounts to
the decomposition of $h$,
\begin{equation}\label{eq:decomposition}
h_{\mu\nu} = h^T_{\mu\nu} + h^{LT}_{\mu\nu} + h^{LL}_{\mu\nu}
+ h^{Tr}_{\mu\nu}\,.
\end{equation}
Here $\,h^{Tr}_{\mu\nu}\,$ is the trace part of $h_{\mu\nu}$ and the
first three terms $\,h^T_{\mu\nu} + h^{LT}_{\mu\nu} +
h^{LL}_{\mu\nu}\,$ comprise its traceless component.  We have the
following well-known identities
\begin{eqnarray}\nonumber 
&&h^T_{\mu\nu} = \bar{\nabla}_{\mu}\xi_{\nu} + \bar{\nabla}_{\nu} \xi_{\mu} \,,\\\nonumber 
&&h^{LT}_{\mu\nu} = \big(\,\bar{\nabla}_{\mu}\bar{\nabla}_{\nu} - 
\frac{1}{d}\,\bar{g}_{\mu\nu}\bar{\Delta}\,\big)\,\sigma \,,  \\
&&h^{LL}_{\mu\nu} = \frac{1}{d}\,\bar{g}_{\mu\nu}\varphi \,,
\end{eqnarray}
where $\xi_{\mu}$ is a transverse vector field and $\sigma$ and
$\varphi$ are scalar fields. The tensor fields, appearing in this
decomposition, obey the following relations, 
\begin{equation}\label{eq:components}
  \bar{g}^{\mu\nu}h^T_{\mu\nu}=0\,,\quad\bar{\nabla}^{\mu}h^T_{\mu\nu}=0\,,\quad 
  \bar{\nabla}^{\mu}\xi_{\mu}=0\, , \quad \varphi=\bar{g}^{\mu\nu}h_{\mu\nu} \,.     
\end{equation}
The scalar field can be further split into two parts
$\varphi=\varphi_0+\varphi_1$ with $\varphi_0$ being orthogonal to
$\varphi_1$ and $\hat\sigma$, for the details we refer the reader to
the literature.

Additionally we decompose the ghost as follows 
\begin{equation}\label{eq:ghost}
  C^{\mu} = C^{T,\mu} + \bar{\nabla}^{\mu}\rho  \,,\qquad \qquad\bar{C}_{\mu} 
  = \bar{C}^T_{\mu} + \bar{\nabla}_{\mu}\bar{\rho} \,,
\end{equation}
where $\bar{C}^T_{\mu}$ and $C^{T,\mu}$ are the transverse components
of $\bar{C}_{\mu}$ and $C^{\mu}$, i.e.
$\bar{\nabla}^{\mu}\bar{C}^T_{\mu} = 0$ and $\bar{\nabla}_{\mu}
C^{T,\mu} = 0$, and $\bar{\rho}$,$\,\rho$ are scalar fields.

\section{Graviton two-point function}\label{app:hatG2+Sgf2}
For the computation of the geometrical flows we need the second
covariant derivative of the diffeomorphism-invariant effective action
in combination with the second derivative of the gauge fixing term,
$\nabla_\gamma^2 \hat \Gamma_{\tiny\mbox{EH}}+ S^{(2)}_{\rm gf}$, see
\eq{eq:approxprop}. For this purpose we need the correction to the
second derivative related to the Riemannian connection
$\Gamma_\gamma$, see e.g.\ \cite{DeWitt:2003pm}. It is given by 
\begin{widetext}
\begin{eqnarray}\nonumber 
 &&\int d^dx'' \left( 
    \Gamma_{\gamma}^{\mu\nu}{}^{\rho'\sigma'}_{\lambda''\tau''}(x,x',x'') 
    \frac{\hat\Gamma_{\tiny\mbox{EH}}[\bar{g}]}{\delta
      \bar{g}_{\lambda''\tau''}(x'')}\right) \\\nonumber 
=&&
  2\kappa^2Z_{N,k}\,\delta(x-x')\sqrt{\bar{g}(x)}
  \sqrt{\bar{g}(x')}\times \Bigg[\Big(\bar{g}^{\mu\rho'}
  \bar{g}^{\sigma'\nu}+\bar{g}^{\mu\sigma'}\bar{g}^{\rho'\nu}-
  \bar{g}^{\mu\nu}\bar{g}^{\rho'\sigma'}\Big)\left(\frac{2+d+
      2\theta d}{8(2+\theta d)}\bar{R}\,
    -\frac{4+d
      +2\theta d}{4(2+\theta d)}\Lambda_k\right)\\
&&+\frac{1}{4}\,\Big(\,\bar{g}^{\mu\nu} 
  \bar{R}^{\rho'\sigma'}+\bar{g}^{\rho'\sigma'}
\bar{R}^{\mu\nu}\Big)-\frac{1}{4}\,\Big(\,\bar{g}^{\nu\rho'}
  \bar{R}^{\sigma'\mu}+\bar{g}^{\nu\sigma'}
  \bar{R}^{\rho'\mu}+\bar{g}^{\mu\rho'}\bar{R}^{\sigma'\nu}
  +\bar{g}^{\mu\sigma'}\bar{R}^{\rho'\nu}\Big)\Bigg]\,, 
\label{eq:Vilcor}\end{eqnarray}
computed at vanishing ghost fields. With \eq{eq:Vilcor} we arrive at  
\begin{eqnarray}\nonumber 
  \int d^d x\,d^d x'\,  h(x)\cdot \left(\nabla_\gamma^2 
    \hat\Gamma_{\tiny\mbox{EH}}[\bar{g}]+
    \mathbf{S}^{(2)}_{\rm gf}[\bar{g}]\right)(x,x')\cdot  h(x')&=&2\kappa^2Z_{N,k}\int 
  d^dx\,\sqrt{\bar{g}}\,h_{\mu\nu}\Bigg[-\Bigg(\frac{1}{2}
  \delta^{\mu}_{\rho'}\delta^{\nu}_{\sigma'}+
  \frac{\theta^2-2\alpha}{4\alpha}\,\bar{g}^{\mu\nu}
  \bar{g}_{\rho'\sigma'}\,\Bigg)\,\bar{\Delta}\end{eqnarray}
\begin{eqnarray}
  &&\hspace{-3cm}+\frac{2-d}{8(2+\theta d)}\Big(2\delta^{\mu}_{\rho'}
  \delta^{\nu}_{\sigma'}\,-\bar{g}^{\mu\nu}\bar{g}_{\rho'\sigma'}\Big)\bar{R}
  -\Bigg(\,\frac{1}{2}-\frac{4+d+2\theta d}{4(2+
    \theta d)}\Bigg)\Big(\,2\delta^{\mu}_{\rho'}\delta^{\nu}_{\sigma'}
  -\bar{g}^{\mu\nu}\bar{g}_{\rho'\sigma'}\Big)\Lambda_k\\\nonumber 
  &&\hspace{-3cm}+\frac{1}{2}\,\bar{g}^{\mu\nu}\,\bar{R}_{\rho'\sigma'}-
  \bar{R}^{\nu}{}_{\rho'}{}^{\mu}{}_{\sigma'}-\frac{\theta+\alpha}{
    \alpha}\big(\,\bar{g}^{\mu\nu}\bar{\nabla}_{\rho'}\bar{\nabla}_{\sigma'}
  \big)+\frac{1-\alpha}{\alpha}\big(
  -\delta^{\mu}_{\sigma'}\bar{\nabla}^{\nu}\bar{\nabla}_{\rho'}
  \big)\Bigg]h^{\rho'\sigma'}\,, 
\label{eq:hatG2+Sgf2}\end{eqnarray}
where $\bar\Delta=\Delta_{\bar g}$. Inserting the York decomposition
detailed in Appendix~\ref{app:york} in \eq{eq:hatG2+Sgf2} finally
leads to
\begin{eqnarray}
&&\hspace{-2.5cm} \int d^d x\,d^d x'\,  h(x)\cdot \left(\nabla_\gamma^2 
    \hat\Gamma_{\tiny\mbox{EH}}[\bar{g}]+
    \mathbf{S}^{(2)}_{\rm gf}[\bar{g}]\right)(x,x')\cdot  h(x')=
 \kappa^2 Z_{N,k} \int d^dx \sqrt{\bar{g}}\\\nonumber 
& &\times \Bigg[ 
 h^T_{\mu\nu} \Biggl[ - 
\bar{\Delta} + A_T(d,\theta)\bar{R}+H_T(d,\theta)\Lambda_k 
\Biggr]h^{T,\mu\nu}\\\nonumber 
  &&+\frac{2}{\alpha}\,\hat{\xi}_{\mu} \Biggl[\left(-\bar{\Delta}
-\frac{\bar{R}}{d}  \right)\left(-\bar{\Delta}+A_V(d,\alpha,\theta)
\bar{R}+H_V(d,\alpha,\theta)\Lambda_k\right)\Biggr]\hat{\xi}^{\mu}\\\nonumber 
  &&+
C_{S2}(d,\alpha)\hat{\sigma} \Biggl[\Bigl(-\bar{\Delta}+A_{S2}(d,\alpha,\theta)\bar{R}+B_{S2}(d,
\alpha,\theta)\Lambda_k \Bigr) \bar{\Delta}\left(\bar{\Delta}
 + \frac{\bar{R}}{d-1}\right) \Biggr]\hat{\sigma} \\ \nonumber 
  &&+ 2C_{S2}(d,\alpha)
C_{S3}(d,\alpha,\theta)\varphi\Bigg[\bar{\Delta}\Bigg(\bar{\Delta} 
+\frac{\bar{R}}{d-1}\,\Bigg)\Bigg]\hat{\sigma}\\ \nonumber 
  &&+C_{S2}(d,\alpha)
C_{S1}(d,\alpha,\theta)\, \varphi \Big[
-\bar{\Delta} + A_{S1}(d,\alpha,\theta)\bar{R} + 
B_{S1}(d,\alpha,\theta)\Lambda_k  \Big]\varphi \Bigg]\,, 
\label{eq:yorkhatG2+Sgf2}\end{eqnarray}
\end{widetext}
where each line in \eq{eq:yorkhatG2+Sgf2} contains the kinetic
operator of the respective field modes. Note in this context that
$\varphi=\varphi_0+\varphi_1$ with $\varphi_0$ being orthogonal to $\varphi_1$ and 
$\hat\sigma$. 

The coefficients of the transversal $h^T$-mode are 
\begin{eqnarray}\nonumber 
  A_T(d,\alpha)&=& \frac{d(d-1)(2-d)+4(2+
\theta d)}{2d(d-1)(2+\theta d)} \,,\\
H_T(d,\theta)&=&\frac{d}{2+\theta d}\,,
\label{eq:ht}\end{eqnarray}
that of the longitudinal mode $\hat\xi$ are 
\begin{eqnarray}\nonumber 
  A_V(d,\alpha,\theta)&=&\frac{\alpha d(2-d)
-2(2+\theta d)}{2d(2+\theta d)}\,, \\\nonumber 
H_V(\alpha)  &=& -2\alpha\,. \label{eq:hatxi}
\end{eqnarray}
The scalar $\hat\sigma,\varphi$-modes have the curvature-coefficients
 \begin{eqnarray}\nonumber 
  A_{S1}(d,\alpha,\theta)&=& \\\nonumber 
&&\hspace{-2.3cm}-\frac{\alpha(d-2)(d^2
-2d+8+4\theta d)}{2(2+\theta d)\left[2
\alpha(d-1)(d-2)-(\theta^2 d^2-4d-4\theta)\,\right]}\\\nonumber 
A_{S2}(d,\alpha,\theta)&=&\frac{\alpha d(2-d)-4(2+
\theta d)}{2(2+\theta d)\left[\,2(d-1)-\alpha(d-2)\,\right]}\,,
\label{eq:curvature-hatsigmavarphi}
\end{eqnarray}
and the coefficients of the cosmological constant terms
 \begin{eqnarray}\nonumber 
B_{S1}(d,\alpha,\theta )&=&\\\nonumber 
&&\hspace{-2.3cm}-\frac{\alpha d^2(2-d)}{(2+
\theta d)\left[\,2\alpha(d-1)(d-2)-(\theta^2 d^2
-4d-4\theta)\,\right]}\,, \\\nonumber 
B_{S2}(d,\alpha,\theta )&=&\frac{\alpha d^2}{(
2+\theta d)\left[\,2(d-1)-\alpha(d-2)\,\right]}\,. 
\label{eq:hatsigmavarphi}
\end{eqnarray}
The scalar terms also have the overall prefactors 
\begin{eqnarray}\nonumber 
C_{S1}(d,\alpha,\theta)&=&\frac{2\alpha(d-1)(2-d)
+(\theta^2 d^2-4d-4\theta)}{2(d-1)\left[\,2(d-1)
-\alpha(d-2)\,\right]}\,,\\\nonumber 
C_{S2}(d,\alpha)&=&\frac{d-1}{d^2}\,\frac{2(d
-1)-\alpha(d-2)}{\alpha}\,, \\ 
C_{S3}(d,\alpha,\theta)&=&\frac{d(-\theta-\alpha)-
2(1-\alpha)}{2(d-1)-\alpha(d-2)}  \,.  
\label{eq:Cprefactors}
\end{eqnarray}
Particularly interesting for the regulators are the coefficients and
prefactor of the kinetic operator $\Delta_{\bar g}$, see 
Appendix~\ref{app:regulators}. 

\section{Ghost two-point function}\label{app:ghostG2}

As for the graviton we split the ghost into its transverse and
longitudinal components and put $g=\bar g$.  In a slight abuse of
notation we write
\begin{equation}\label{eq:ghostT+L}
C^{\mu} = C^{T,\mu} + \bar\nabla^{\mu}\0{1}{\sqrt{-\bar\Delta} }\eta\,,\quad 
\bar C_{\mu} = \bar C^T_{\mu} + \bar\nabla_{\mu}\0{1}{ \sqrt{-\bar\Delta} }\bar\eta \,,
\end{equation}
neglecting the subtleties concerning the inversion of $\bar\Delta$.
In \eq{eq:ghostT+L} $\bar{C}^T_{\mu}$ and $C^{T,\mu}$ are the transverse components
of $\bar C_{\mu}$ and $C^{\mu}$, i.e. $\bar{\nabla}^{\mu}\bar
C^T_{\mu} = 0$ and $\bar\nabla_{\mu} C^{T,\mu} = 0$, and $\eta$,
$\bar\eta$ are scalar Grassmann fields. Inserting the parameterisation \eq{eq:ghostT+L} 
in the ghost action, \eq{eq:finalS},  we finally arrive at 
\begin{eqnarray}\nonumber 
S_{gh} &=&2 \int d^d x\sqrt{\bar g }\,\bar{C}^T_{\mu} \left(-\bar\Delta -
 \0{\bar R}{d} \right) C^{T,\mu} \\
& &+ 2\int d^d x\sqrt{\bar g }\,\bar\eta\left(-\bar\Delta 
-\0{2\bar R}{d} \right)\eta\,. 
\label{eq:ghostsplit}\end{eqnarray}

\section{Regulators}\label{app:regulators}

The following appendix contains the full set of regulators for the
flow within the background field approximation, see
Section~\ref{sec:backgroundflow}. Our choice is adapted to the York
transverse-traceless decomposition of the kinetic term as detailed in
the last Appendix~\ref{app:hatG2+Sgf2}. The full regulator is chosen
diagonal in the basis in field space provided by the York
decomposition. Below we provide the scalar parts of the regulators,
the lower indices refer to the modes in field space. With
\begin{equation} 
\bar x=-\0{\Delta_{\bar g}}{k^2}\,, 
\end{equation} 
the regulator for a general mode of the York decomposition simply
amounts to 
\begin{eqnarray}\label{eq:kin+regs}
-\bar\Delta \to -\bar\Delta + k^2\,r(\bar x)\,,
\end{eqnarray}
for the terms proportional to $\bar\Delta$ in \eq{eq:yorkhatG2+Sgf2}. 
This choice respects the diagonality of the York decomposition. For example, 
the kinetic operator on the $h^T$-subspace reads $-Z_{N,k}
\kappa^2\bar\Delta$ and hence we choose the regulator 
\begin{equation}
(R_k [\bar{g}])_{h^Th^T} = Z_{N,k}  \kappa^2  k^2 r(\bar x) \,, 
\end{equation}
where we have dropped the projection operator on the $h^T$-subspace.  The
regulators on the $h^{TL}$ subspace are given by
\begin{eqnarray}\nonumber 
&& \hspace{-1cm}\big(R_k [\bar{g}]\big)_{\varphi_1\sigma}=  Z_{N,k} \, C_{S2}(d,\alpha)
C_{S3}(d,\alpha) \kappa^2 \\\nonumber 
&   & k^2 \Bigg( - \sqrt{\bar x \left(\bar x - \frac{ \bar R/k^2 }{d-1} \right) }\\\nonumber 
& & +\sqrt{\bar x -\frac{\bar R/k^2}{d-1} +  r(\bar x)} 
\sqrt{\bar x + r(\bar x) }\Bigg)\,, 
\end{eqnarray}
with
$(R_k)_{\varphi_1\sigma}=(R_k[\bar{g})_{\sigma\varphi_1}^{\dag}$, as well as 
\begin{eqnarray}\nonumber 
 (R_k[\bar{g}])_{\sigma\sigma} &=&Z_{N,k}\, C_{S2}(d,\alpha)
  \kappa^2 k^2 r(\bar x)\,, 
\end{eqnarray}
and for $i=0,1$, 
\begin{eqnarray}\nonumber 
  (R_k [\bar{g}])_{\varphi_i\varphi_i}
&=&Z_{N,k}\, C_{S2}(d,\alpha)
  C_{S1}(d,\alpha)\kappa^2 k^2 r(\bar x)\,.\hspace{.5cm} 
\end{eqnarray}
The regulator on the $h^T_{\mu\nu} \times h^T_{\mu\nu}$-subspace is
given by 
\begin{eqnarray} 
  \big(R_k [\bar{g}]\big)_{\xi\xi} = Z_{N,k} \frac{2}{\alpha}\,\kappa^2  k^2 r(\bar x)\,, 
\end{eqnarray}
where again we dropped the projection operator. 
Finally, the regulators of the ghost modes are given by 
\begin{eqnarray}
(R_k[\bar{g}])_{\bar{C}^T C^T}& =&   
2\,k^2 r(\bar x)\nonumber\\
(R^{gh}_k[\bar{g}])_{\bar{\eta} \eta} &=& 2 k^2 r(\bar x)\,,
\end{eqnarray}
where we have $(R_k)_{\bar{C}^T C^T}=-(R_k)_{ C^T \bar C^T}$ and 
$(R_k)_{\bar\eta \eta}=-(R_k)_{ \eta \bar \eta}$.

\section{Threshold functions and coefficient functions $\bar F_\lambda$ and
  $\bar F_g$}\label{app:Is}
The loop integrals in the flow equations for the couplings are
represented by the coefficient functions $I$, which are, up to
prefactors, the standard threshold functions. In the present case
these threshold functions only depend on the constant part $c_\lambda
\lambda$ of the two-point functions. For general regulators
\eq{eq:regulators} with shape function $r$ the threshold functions
read
\begin{eqnarray}\nonumber 
  \Phi^p_n(\omega)&=& \0{1}{\Gamma(n)}
  \int_0^{\infty} dx\,x^{n-1} \frac{r(x)-x r'(x)}{(x+r(x)+
\frac{d}{d-2}\omega)^p} \,, \quad \\[1ex] 
  \tilde \Phi^p_n(\omega)&=&\0{1}{\Gamma(n)}
  \int_0^{\infty} dx\,x^{n-1} \frac{r(x)}{(x+r(x)+
\frac{d}{d-2}\omega)^p} \,, 
\label{eq:thres} 
\end{eqnarray}
where $\omega$ is $0$ or $-\lambda$, depending on the mode
considered. The threshold function $\Phi^p_n$ appears in terms
proportional to $\partial _t r(x)$ leading to the coefficient
functions $I^{(1)}$ in the flow equations.  The threshold function
$\tilde \Phi^p_n$ appears in terms proportional to $\partial_t Z$ or
$\partial_t \lambda$, leading to the coefficient functions $I^{(2)}$.

For the flow of the cosmological constants $\bar\lambda$ and
$\lambda$, \eq{eq:flowbackLa} and \eq{eq:flowLafluc} respectively, the
coefficient functions read  
\begin{eqnarray}\nonumber 
\bar F_\lambda^{(1)}  &=& \0{ 4 \pi g_N }{(4\pi)^{d/2}} \left[d(d-1)
  \Phi^1_{\frac{d}{2}}(-\lambda)-d\, \Phi^1_{\frac{d}{2}}(0)\right]\,, \\[1ex]
\bar F^{(2)}_{\lambda}  &=&  \0{1}{2} \0{4 \pi g_N}{(4\pi)^{d/2}} d(d-1)
\tilde{\Phi}^1_{\frac{d}{2}} (-\lambda)\,. 
\label{eq:flowLaI}
\end{eqnarray}
The last coefficient function is that of the ghost loop,
$I^{(1)}_{\lambda,\rm gh}=I_{\lambda,\rm gh}$, there is no term
proportional to the wave function renormalisation of the ghost which
we dropped in the present analysis. 

The coefficient functions in the flow of the Newton constant
$\partial_t \bar g_N$ and $\partial_t g_N$, \eq{eq:flowbackZN} and
\eq{eq:flowZNfluc} respectively, read 
\begin{eqnarray}\nonumber 
\bar F_g^{(1)}& =&  4 \0{4 \pi g_N }{(4\pi)^{d/2}}
\left[ a_1\Phi^1_{\frac{d-2}{2}} (-\lambda) 
    +a_2\Phi^2_{\frac{d}{2}} (- \lambda)
  \right]\\[1ex]\nonumber 
&&- 4\0{4 \pi g_N }{(4\pi)^{d/2}}
\left[a_3 \Phi^1_{\frac{d-2}{2}}(0)+ 
    a_4 \Phi^2_{\frac{d}{2}} (0)\right]\,,\\[1ex]
\bar F_g ^{(2)}& =&  2 \0{4 \pi g_N}{(4\pi)^{d/2}}
\left[a_1 \tilde{\Phi}^1_{\frac{d-2}{2}}(-
      \lambda) +
      a_2 \tilde{\Phi}^2_{\frac{d}{2}} (- \lambda)
     \right]  \,,
\label{eq:flowZNI} 
\end{eqnarray}
with the parameters $a_i$,
\begin{eqnarray} \label{eq:a12}\nonumber 
 a_1 &=& \0{d^3-2 d^2-11 d-12}{12 (d-1)}\,, \\[1ex] 
 a_2 & =&  -\0{d^3-2 d^2+5  d-12 }{4 (d-1)} \,,
\end{eqnarray} 
and 
\begin{eqnarray} \label{eq:a34}
a_3 =\0{d^2-6}{6d} \,, &\quad  & a_4 =\0{d+1}{d}\,, 
\end{eqnarray}
Note that parts of $\bar F_g$ in \eq{eq:flowZNI} stemming from modes
without and with explicit curvature-dependence are those proportional
to $\Phi^1_{\frac{d-2}{2}},\tilde\Phi_{\frac{d-2}{2}}^1$, and
$\Phi^2_{\frac{d}{2}},\tilde\Phi_{\frac{d}{2}}^2$,
respectively. Hence, it is the index $p-1=0,1$ of $\Phi^p,
\tilde\Phi^p$ which labels the modes without, $p-1=0$, and with,
$p-1=1$, curvature-dependence.

\section{Threshold functions and coefficient functions $F_\lambda$ and
  $F_g$ for the optimised regulator}\label{app:thresopt}
For the optimised regulator the threshold functions \eq{eq:thres} in Appendix A read, 
\begin{eqnarray}\label{eq:optthreshold} 
&&\Phi^p_n(\omega)  =  \frac{1}{n\,\Gamma(n)}\,\frac{1}{(1+\frac{d}{d-2}\omega)^p}\,, \\
&&\tilde{\Phi}^p_n(\omega)  =  \frac{1}{n+1}\Phi^p_n(\omega) \,.\label{eq:optrelgen} 
\end{eqnarray} 
\Eq{eq:optthreshold} implies that 
\begin{eqnarray}\label{eq:optrel}
\tilde \Phi^p_{\frac{d-2}{2}}(\omega)= \0{2}{d}\Phi^p_{\frac{d-2}{2}}(\omega)\,,
\quad \tilde \Phi^p_{\frac{d}{2}}(\omega)= \0{2}{d+2}\Phi^p_{\frac{d}{2}}(\omega)\,.
\end{eqnarray}
Inserting these relations into \eq{eq:flowLaI}, \eq{eq:flowZNI}, leads
to \eq{eq:Iconstraint} with the subscript $p-1=0,1$ labels the
curvature-dependence. Then the $\bar F_{\lambda}$'s, \eq{eq:flowLaI}, used
in Section~\ref{sec:dynflow} for the flow of the background 
dynamical cosmological constant,  are given by
\begin{eqnarray} \nonumber 
  \bar F_{\lambda}^{(1)} &= & 6 \frac{g_N}{4\pi} \, \left(-\frac{2}{3}+\frac{1}{1-2\lambda} \right) \,,\\[1ex]
 \bar F_{\lambda}^{(2)} &=&  \frac{g_N}{4\pi} \, \frac{1}{1-2\lambda} \,. 
\label{eq:Flaopt}\end{eqnarray}
\Eq{eq:Ilaopt} leads to the coefficients $I_\lambda$, 
\begin{eqnarray}\nonumber 
I_{\lambda,-2}& =& \frac{g_N}{4\pi} \, \frac{1}{1-2\lambda}  \\
I_{\lambda,\rm gh} & =& -  \0{g_N}{4 \pi}\frac{4}{3}\,, 
\label{eq:Ilaopt}\end{eqnarray} 
with $I_{\lambda,\rm gh}=-2I_{\lambda,0}$.  The $\bar F_g$'s,
\eq{eq:flowZNI}, used in Section~\ref{sec:dynflow} for the flow of the
background cosmological constant, \eq{eq:flowbackZNopt}, are given by
\begin{eqnarray} \nonumber 
\bar F_{g}^{(1)}& =& -4 \0{g_N^2}{4\pi} \,\left(\frac{25}{24}+\frac{2}{3}\frac{1}{1-2\lambda} + 
    \frac{5}{3}\frac{1}{(1-2\lambda)^2} \right)  \,, \\[1ex]  
 \bar F_g^{(2)}& =& - \frac{g_N^2}{6\pi} \,\left(\frac{1}{1-2\lambda} + 
   \frac{5}{3}\,\frac{1}{(1-2\lambda)^2} \right)\,. 
\label{eq:FZNopt}\end{eqnarray}
\Eq{eq:IZNopt} leads to the coefficients $I_N$, 
\begin{eqnarray} \nonumber 
 I_{N,-2}& =& - \frac{g_N^2}{6\pi} \,\left(\frac{1}{1-2\lambda} + 
   \frac{5}{3}\,\frac{1}{(1-2\lambda)^2} \right)\,, \\[1ex] \nonumber 
I_{N,\rm gh}& =&   -\0{4}{3}\0{g_N}{3 \pi}  \0{25}{24}\,, \\
I_{N,\rm gh,1}& =&   -\0{1}{3}\0{g_N}{3 \pi}  \0{25}{24}\,, 
\label{eq:IZNopt}\end{eqnarray}
with $I_{N,\rm gh}=-2I_{N,0}$. For the ghost coefficients we have used
that the term independent of $\lambda$ in first line of \eq{eq:FZNopt}
stems from the sum of ghost $I_{N,\rm gh}+I_{N,0}= 1/2 I_{N,\rm
  gh}$. We get with \eq{eq:Isplit} and \eq{eq:optrel} with $d=4$ that 
\begin{equation} \label{eq:extractghost}
-4 \0{g_N^2}{4\pi} \frac{25}{24} = 2 I_{N,\rm gh,0} + 3 I_{N,\rm gh,1}\,,\qquad 
 I_{N,\rm gh,1} =\0{1}{3} I_{N,\rm gh,0}\,.
\end{equation} 
This leads to the $I_{\rm ghost}$ in \eq{eq:IZNopt}. 
The coefficients $I_{N,-2,0}$ and
$I_{N,-2,1}$ are given by the terms in the first line of
\eq{eq:IZNopt} which are proportional to $(1-2 \lambda)^{-1}$ and
$(1-2 \lambda)^{-2}$ respectively.

For the computation of the flow of the dynamical couplings we simply
have to insert the coefficients $I_\lambda$ and $I_N$ in 
\eq{eq:Ilaopt},\eq{eq:IZNopt},\eq{eq:Is} in \eq{eq:flowZNfluc},\eq{eq:flowLafluc}. 
This leads us to the following expressions for the right hand sides $F_g,F_\lambda$ 
of the dynamical flows  \eq{eq:fullflow}, 
\begin{eqnarray}\nonumber 
  F_g(g_N,\lambda) &=&  -\0{g_N^2}{3 \pi} \left(\frac{5}{3}\frac{25}{24}+
    \0{1}{1-2\lambda}+\frac{5}{3}\,\frac{1}{(1-2 \lambda)^2}
  \right) \\[1ex]  
  & &\hspace{-1.5cm}+\left(\eta_N+\dot \lambda\partial_\lambda\right)  
 \0{g^2_N}{6\pi} \left(\0{1}{1-2\lambda}+
    \frac{5}{3}\0{1}{(1-2 \lambda)^2}\right)\,\,,
\label{eq:finalflowFg}\end{eqnarray} 
and 
\begin{eqnarray}\nonumber 
 F_{\lambda}(g_N,\lambda) &=&  - \,\frac{g_N}{2\pi}\left(\0{2}{3}
-\0{1}{1-2\lambda}\right)\\[1ex]
&& -\left(\eta_N+
\0{2 \partial_t \lambda}{1-2 \lambda}\right)
\0{g_N}{4\pi}\0{1}{1-2\lambda}\,.
\label{eq:finalflowFl}
\end{eqnarray}

\end{appendix}
\eject


\end{document}